%% file: trpms-3gamma.tex
\documentclass[journal]{IEEEtran}

\input{./include/preamble}

\input{./include/acronyms}

\input{./include/macros}

\begin{document}

\title{Direct3$\upgamma$: a Pipeline for Direct Three-gamma PET Image Reconstruction
}
\author{
    Youness Mellak, Alexandre Bousse, Thibaut Merlin,  Debora Giovagnoli, Dimitris Visvikis
	\thanks{This work did not involve human subjects or animals in its research.}
	\thanks{All authors declare that they have no known conflicts of interest in terms of competing financial interests or personal relationships that could have an influence or are relevant to the work reported in this paper.}
	\thanks{This work has received a French government support granted  to the Cominlabs excellence laboratory and managed by the French National Research Agency (ANR) in the ``Investing for the Future'' program under reference ANR-10-LABX-07-01.}
	\thanks{All authors  are affiliated to the LaTIM, Inserm, UMR 1101, \emph{Universit\'e de Bretagne Occidentale}, Brest, France.} \thanks{Corresponding author: A. Bousse, \texttt{bousse@univ-brest.fr}}
	}

\maketitle

\input{./content/abstract}

\input{./content/intro}
\input{./content/method}

\input{./content/experiments}
\input{./content/results}

\input{./content/discussion}

\input{./content/conclusion}
\appendices
\input{./content/appendix}

\AtNextBibliography{\footnotesize} 
\printbibliography

\end{document}

%% file: include/preamble.tex
\usepackage{times}

\usepackage{graphicx}
\usepackage{textcomp}

\usepackage{tikz}
\usepackage{tikz-network}
\usepackage{amsmath}
\usepackage{amssymb}
\usepackage{amsfonts}
\usepackage{bm}

\usepackage{acro}[=v2] 
\acsetup{single}

\usepackage{tikz}
\usetikzlibrary{spy}

\usepackage{subcaption}
\usepackage{enumitem}   

\usepackage{overpic}
\usepackage{kbordermatrix}

\usepackage{textcomp}

\usepackage{upgreek}

\usepackage{multirow}
\usepackage{booktabs}
\usepackage{tabularx}
\usepackage{pgfplots}
\pgfplotsset{compat=1.13}
\usetikzlibrary{arrows.meta}
\usepackage[labelformat=simple]{subcaption}

\usepackage[
	style=ieee,
	dashed=false,
	maxbibnames=20,
	minbibnames=2,
	maxcitenames=1,
	mincitenames=1,
	backend=biber,
	defernumbers=false,
	sorting=none,
	useprefix=false
	]{biblatex}

\definecolor{colorBestTOF}{HTML}{3423A6}
\definecolor{colorAIVarSigma}{HTML}{27FB6B}
\definecolor{colorTOF}{HTML}{FB62F6}
\definecolor{colorTOF_100}{HTML}{FB62F6}
\definecolor{colorTOF_200}{HTML}{426B69}
\definecolor{colorAI2}{HTML}{27FB6B}
\definecolor{colorPhys}{HTML}{886F68}
\definecolor{color2Gamma}{HTML}{F5A65B}

\definecolor{colorPhysic}{HTML}{d4abdc}
\definecolor{colorFillPhysic}{HTML}{efd8f4}
\definecolor{colorFCNN}{HTML}{543D46}
\definecolor{colorFillFCNN}{HTML}{B295A0}
\definecolor{colorAI}{HTML}{477d7f}
\definecolor{colorFillAI}{HTML}{64bd97}

\usepackage{tikz}
\usetikzlibrary{arrows,shapes,calc,decorations.pathreplacing,intersections,quotes,angles,positioning,fit,petri,through,shadows,plotmarks,spy,math}

\usepackage{xcolor}
\colorlet{myred}{red!80!black}
\colorlet{myblue}{blue!80!black}
\colorlet{mygreen}{green!60!black}
\colorlet{myorange}{orange!70!red!60!black}
\colorlet{mydarkred}{red!30!black}
\colorlet{mydarkblue}{blue!40!black}
\colorlet{mydarkgreen}{green!30!black}

\colorlet{c1}{myblue!80}
\colorlet{c11}{c1!30}

\colorlet{c2}{myred!80}
\colorlet{c22}{c2!30}

\colorlet{c3}{mygreen!80}
\colorlet{c33}{c3!30}


%% file: include/acronyms.tex
\DeclareAcronym{PET}{
	short=PET,
	long=positron emission tomography,
}

\DeclareAcronym{PINN}{
	short=NN,
	long= physics-informed neural network,
}

\DeclareAcronym{PGNN}{
	short=PGNN,
	long= physics-guided neural network,
}

\DeclareAcronym{NN}{
	short=NN,
	long=neural network,
}

\DeclareAcronym{GNN}{
	short=GNN,
	long=graph neural network,
}

\DeclareAcronym{GRU}{
	short=GRU,
	long=gated recurrent unit,
}

\DeclareAcronym{GCN}{
	short=GCN,
	long=graph convolutional network,
}

\DeclareAcronym{ANN}{
	short=ANN,
	long= artificial neural network,
}

\DeclareAcronym{CNN}{
	short=CNN,
	long=convolutional neural network,
}

\DeclareAcronym{RNN}{
	short=RNN,
	long=recurrent neural network,
}

\DeclareAcronym{FCNN}{
	short=FCNN,
	long=fully-connected neural network,
}

\DeclareAcronym{MLP}{
	short=MLP,
	long=multilayer perceptron,
}	

\DeclareAcronym{LOR}{
	short=LOR,
	long=line of response,
	long-plural-form = lines of response,
}	

\DeclareAcronym{TOF}{
	short=TOF,
	long=time-of-flight
}

\DeclareAcronym{LM}{
	short=LM,
	long=list-mode
}

\DeclareAcronym{AI}{
	short=AI,
	long=artificial intelligence
}

\DeclareAcronym{MPNN}{
	short=MPNN,
	long=message passing neural network
}

\DeclareAcronym{FOV}{
	short=FOV,
	long=field of view,
	long-plural-form = fields of view,
}

\DeclareAcronym{IN}{
	short=IN,
	long=interaction network,
}

\DeclareAcronym{MIN}{
	short=MIN,
	long=modified interaction network,
}

\DeclareAcronym{MDN}{
	short=MDN,
	long=mixture density network,
}

\DeclareAcronym{2D}{
	short=2-D,
	long=two-dimensional,
}

\DeclareAcronym{3D}{
	short=3-D,
	long=three-dimensional,
}

\DeclareAcronym{1D}{
	short=1-D,
	long=one-dimensional,
}

\DeclareAcronym{GATE}{
	short=GATE,
	long=Geant4 Application for Tomography Emission,
}

\DeclareAcronym{PDF}{
	short=PDF,
	long=probability density function,
}

\DeclareAcronym{FWHM}{
	short=FWHM,
	long=full width at half maximum ,
}

\DeclareAcronym{RMSE}{
	short=RMSE,
	long=root-mean-square error,
}

\DeclareAcronym{MLEM}{
	short=MLEM,
	long=maximum-likelihood expectation-maximization algorithm,
}

\DeclareAcronym{SNR}{
	short=SNR,
	long=signal-to-noise ratio,
}

\DeclareAcronym{PSNR}{
	short=PSNR,
	long=peak signal-to-noise ratio,
}

\DeclareAcronym{SSIM}{
	short=SSIM,
	long=structural similarity index measure,
}

\DeclareAcronym{MAE}{
	short=MAE,
	long=mean absolute error,
}

\DeclareAcronym{LHC}{
	short=LHC,
	long=large hadron collider,
}

\DeclareAcronym{XCAT}{
	short=XCAT,
	long=Extended Cardiac-Torso,
}

\DeclareAcronym{EM}{
	short=EM,
	long=expectation maximization,
}

\DeclareAcronym{castor}{
	short=CASToR,
	long=the Customizable and Advanced Software for Tomographic Reconstruction,
}

\DeclareAcronym{GT}{
	short=GT,
	long=ground truth,
}

\DeclareAcronym{ROI}{
	short=ROI,
	long=region of interest,
}

\DeclareAcronym{STD}{
	short=STD,
	long=standard deviation,
}

\DeclareAcronym{LXe}{
	short=LXe,
	long=liquid xenon,
}

\DeclareAcronym{DREP}{
    short=DREP,
	long=detector response error propagator}

\DeclareAcronym{DRF}{
    short=DRF,
	long=detector response function 
}

\DeclareAcronym{RSS}{
    short=RSS,
	long=root sum square
}

 \DeclareAcronym{ToM}{
    short=ToM,
	long=tri-oriented Mamba}

 \DeclareAcronym{GSC}{
    short=GSC,
	long=gated spatial convolution}

 \DeclareAcronym{AG}{
    short=AG,
	long=attention gate
}

  \DeclareAcronym{LS}{
    short=LS,
	long=least-square
}

 \DeclareAcronym{GAN}{
    short=GAN,
	long=generative adverserial network
}
 
 \DeclareAcronym{HO}{
    short=HO,
	long=histo-image with ordering algorithm}

 \DeclareAcronym{OO}{
    short=OO,
	long=output with ordering algorithm
}

\DeclareAcronym{H2P}{
    short=H2P,
	long=histo-image with two points,
	long-plural-form = histo-images with two points,
}
 
\DeclareAcronym{O2P}{
    short=O2P,
	long=output with two points,
	long-plural-form = outputs with two points,
}

 \DeclareAcronym{HMC}{
    short=HMC,
	long=histo-image with multiple cones,
	long-plural-form = histo-images with multiple cones,
}

 \DeclareAcronym{OMC}{
    short=OMC,
	long=output with multiple cones,
	long-plural-form = outputs with multiple cones,
}

\DeclareAcronym{PR}{
	short=PR,
	long=positron range,
}

\DeclareAcronym{KN}{
	short=KN,
	long=Klein-Nishina,
}

\DeclareAcronym{3g}{
	short=3-$\upgamma$,
	long=three-gamma,
}

\DeclareAcronym{MC}{
	short=MC,
	long=Monte Carlo,
}

\DeclareAcronym{XEMIS}{
	short=XEMIS2,
	long=Xenon Medical Imaging System 2,
}

\DeclareAcronym{DIP}{
	short=DIP,
	long=deep image prior,
}

\DeclareAcronym{LSO}{
	short=LSO,
	long=lutetium oxyorthosilicate,
}

\DeclareAcronym{TPC}{
	short=TPC,
	long=time projection chamber,
}

\DeclareAcronym{CNR}{
	short=CNR,
	long=contrast-to-noise ratio,
}

\DeclareAcronym{CRC}{
	short=CRC,
	long=contrast-recovery coefficient,
}

\DeclareAcronym{RC}{
	short=RC,
	long=recovery coefficient,
}

%% file: include/macros.tex
\newcommand{\boldb}{\bm{b}}

\newcommand{\bolde}{\bm{e}}

\newcommand{\boldh}{\bm{h}}

\newcommand{\boldn}{\bm{n}}
\newcommand{\boldo}{\bm{o}}
\newcommand{\boldp}{\bm{p}}

\newcommand{\boldr}{\bm{r}}

\newcommand{\boldx}{\bm{x}}
\newcommand{\boldy}{\bm{y}}
\newcommand{\boldz}{\bm{z}}

\newcommand{\boldC}{\bm{C}}

\newcommand{\boldE}{\bm{E}}
\newcommand{\boldF}{\bm{F}}
\newcommand{\boldG}{\bm{G}}

\newcommand{\boldM}{\bm{M}}

\newcommand{\boldO}{\bm{O}}

\newcommand{\boldR}{\bm{R}}

\newcommand{\calE}{\mathcal{E}}

\newcommand{\calL}{\mathcal{L}}

\newcommand{\rmc}{\mathrm{c}}
\newcommand{\rmd}{\mathrm{d}}
\newcommand{\rme}{\mathrm{e}}
\newcommand{\rmf}{\mathrm{f}}

\newcommand{\rmo}{\mathrm{o}}

\newcommand{\rmr}{\mathrm{r}}
\newcommand{\rms}{\mathrm{s}}

\newcommand{\rmE}{\mathrm{E}}

\newcommand{\rmO}{\mathrm{O}}

\newcommand{\rmR}{\mathrm{R}}

\newcommand{\boldtheta}{\bm{\theta}}
\newcommand{\boldpsi}{\bm{\psi}}

\newcommand{\boldvartheta}{\bm{\vartheta}}
\newcommand{\R}{\mathbb{R}}

\newcommand{\transp}{^\top}

\newcommand{\argmin}{\operatornamewithlimits{arg\,min}}

%% file: content/abstract.tex
\begin{abstract}
	\bfseries\boldmath This paper presents a novel image reconstruction pipeline for \ac{3g}  \ac{PET} aimed at improving spatial resolution and reducing noise in nuclear medicine. The proposed Direct3$\upgamma$ pipeline addresses the inherent challenges in \ac{3g} \ac{PET} systems, such as detector imperfections and uncertainty in photon interaction points. A key feature of the pipeline is its ability to determine the order of interactions through a model trained on \ac{MC} simulations using the \ac{GATE} toolkit, thus providing the necessary information to construct Compton cones which intersects with the \ac{LOR} to provide an estimate of the emission point. The pipeline processes \ac{3g} \ac{PET} raw data, reconstructs histoimages by propagating energy and spatial uncertainties along the \ac{LOR}, and applies a \acs{3D} \ac{CNN} to refine these intermediate images into high-quality reconstructions. To further enhance image quality, the pipeline leverages both supervised learning and adversarial losses, the latter preserving fine structural details. Experimental results show that Direct3$\upgamma$ consistently outperforms conventional 200-ps \ac{TOF} \ac{PET} in terms of \ac{SSIM} and \ac{PSNR}. 
\end{abstract}


%% file: content/intro.tex
\section{Introduction}\label{sec:introduction}

\IEEEPARstart{S}{ince} the early 2000's the idea of \ac{3g}  \ac{PET} imaging has been considered. It is based on the utilization of radioisotopes that emit a positron and almost simultaneously an additional gamma photon. Such popular non-pure positron emitters include \textsuperscript{124}I (half-life=4.176d, emission of 602.7~keV photon) and \textsuperscript{44}Sc (half-life=4.176d, emission of 1.157-MeV photon) which have been associated with multiple clinical applications in oncology and, more specifically, in the field of theranostics \cite{kist2016other,muller2013promises}. 

\begin{figure}[hbt]
	\centering
	\includegraphics[width=0.7\columnwidth]{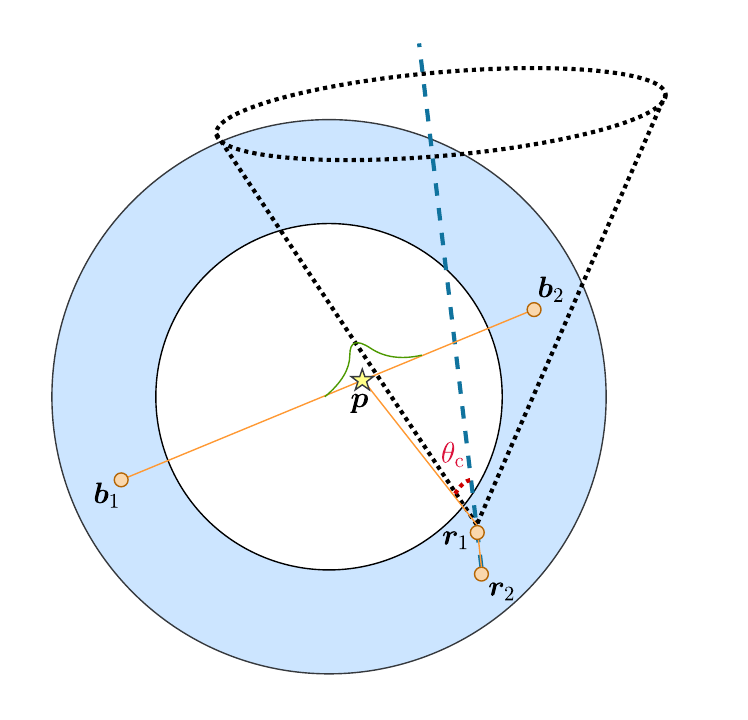}%
	\caption{Estimating the point of emission using Compton kinematics.
		$\bm{b}_1$ and $\bm{b}_2$ are respectively the first and second photon detected positions resulting from the  annihilation (back-to-back photons) while $\bm{r}_1$ and $\bm{r}_2$ are respectively the  first and second interaction positions of the prompt gamma in the detector. The angle $\theta_\rmc$, estimated using \eqref{eq:kn}, is used to draw the Compton cone. The yellow star indicates the real emission point $\boldp$. 
	}
	\label{Cone_LOR}
\end{figure}

The principle of \ac{3g} \ac{PET} imaging is based on the general concept of a Compton camera, where the detection of a Compton scatter and associated kinematics is used for the reconstruction of the source position.  The annihilation of the positron with an electron in the tissues produces the two 511-keV photons defining the coincidence \ac{LOR} is used in conventional \ac{PET} imaging.  Through the emission of the additional (third) gamma, Compton interactions in the detector can be used to define a Compton cone using Compton kinematics. The cone is drawn on the basis of the first two interactions of the third gamma in the detector. More specifically, the aperture angle of the cone is given as the Compton angle of the first interaction, while the axis of the cone is the line connecting the two interactions. The intersection between the cone and the \ac{LOR} can be subsequently used to provide additional information about the position of the source (see Figure~\ref{Cone_LOR}). 

In the past two decades, two different detector systems have been developed for the implementation of such \ac{3g} \ac{PET} imaging systems. First, the \ac{LXe} Compton camera (e.g., the \ac{XEMIS} project \cite{manzano2015xemis}) where the \ac{LXe} acts as the scatterer and detection medium for the third photon, but also for the two back-to-back 511-keV photons \cite{manzano2018xemis2}. The second system involves the utilization of a dual-detector structure combining \ac{PET} and Compton imaging, with the second detector acting as the scatterer \cite{thirolf2015perspectives,yamaya2017whole}.  

Recent advances in \ac{3g} imaging reconstruction techniques have been aimed at improving image quality while using low statistics. \citeauthor{Giovagnoli}~\cite{Giovagnoli}  introduced a method that uses the intersection point of the Compton cone and two coincidence photons of the \ac{LOR} as the center for a \ac{PDF}, similar to \ac{TOF} \ac{PET}. \citeauthor{yoshida2020whole}~\cite{yoshida2020whole} proposed a scanner design with separate scatterer and absorber modules, incorporating scatter angle calculations using the Compton scatter formula and modeling blurring with asymmetric Gaussian functions. Both approaches rely on identifying the \ac{LOR}-Compton cone intersection point, but face challenges in determining the order of prompt gamma interactions. Yoshida's scatter-absorber design addresses this through hardware modifications and energy windowing, albeit at the cost of reduced sensitivity. In contrast, the \ac{XEMIS}-like scanner uses a single dense ring of \ac{LXe}, where prompt gammas interact multiple times until absorption. This design, while promising, lacks a built-in mechanism for determining the interaction order and is susceptible to errors from spatial resolution limitations and Doppler effects, especially given the proximity of interaction points.

\begin{figure}[hbt]
	\centering
	\includegraphics[width=0.485\textwidth]{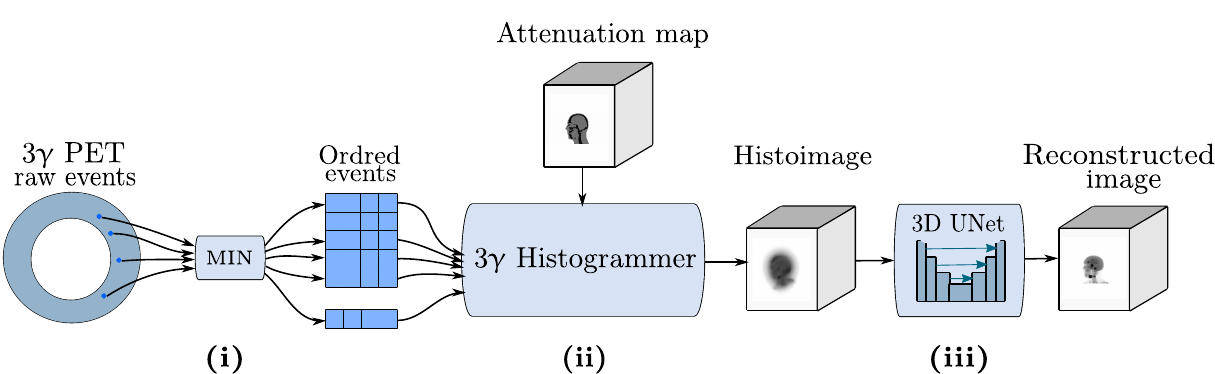}
	\caption{Direct3$\upgamma$ pipeline, from event detection, building histoimage using the histogrammer and the backbone to reconstruct the final image.}
	\label{fig:DIRECT3gammaFinal}
\end{figure}

Our proposed method, Direct3$\upgamma$, presents a structured pipeline for reconstructing \ac{3D}  \ac{PET} images from \ac{3g} data on an event-by-event basis. This approach consists of three main stages, as illustrated in Figure~\ref{fig:DIRECT3gammaFinal}:
\begin{enumerate}[label=(\roman*)]
	\item Event Detection and Compton Cone Construction: We start by detecting \ac{3g} events from the scanner data. Next, we use a \ac{GNN}-based architecture, referred to as \ac{MIN}, to sequence the interactions of photons. The training of \ac{MIN} is achieved on \ac{MC}-simulated data obtained from \ac{GATE} \cite{jan2004gate}. Finally, we construct the Compton cone on the basis of first- and second-interactions.
	\item \Ac{LOR} Processing and Histoimage Generation: This stage involves accounting for blur introduced by the uncertainties in determining the Compton cone using the \ac{DREP} method, applying attenuation correction, and generating a histoimage (i.e., the backprojected list-mode data) as an initial representation of the activity distribution.
	\item Image Reconstruction and Enhancement: The final stage employs an encoder-decoder image processing \ac{CNN}, trained as a \ac{GAN}, performing both deblurring and denoising to enhance image quality.
\end{enumerate}

Various approaches have been proposed to address the challenges in stage (i). \citeauthor{oberlack2000compton}~\cite{oberlack2000compton} presented an algorithm to determine the Compton interaction sequence by minimizing a $\mathrm{d}\phi$-criterion among $N!$ possible sequences. \citeauthor{pratx2009bayesian}~\cite{pratx2009bayesian} introduced a Bayesian approach utilizing additional information about photon interactions and detector characteristics. Although effective, these methods are computationally expensive. \citeauthor{NNComptonKinematicsar}~\cite{NNComptonKinematicsar} developed a \ac{NN} to improve efficiency, but it remains limited for complex scenarios ($N>4$). To overcome these limitations, we propose a new approach inspired by \citeauthor{andersson2021gamma}~\cite{andersson2021gamma}. Our method leverages a \ac{GNN} with a \ac{MIN} to classify edges and determine the path of the prompt gamma in the detector.

For stage (ii), we build on previous work in incorporating detector uncertainties. \citeauthor{Giovagnoli}~\cite{Giovagnoli} propagated spatial and energy uncertainties to angle uncertainties, modeling them as symmetric Gaussian functions on the \ac{LOR}. \citeauthor{yoshida2020whole}~\cite{yoshida2020whole} proposed the use of a nonsymmetric Gaussian function and noted that position estimations are highly accurate when this angle is close to 90\textsuperscript{o}. However, as the angle approaches 0\textsuperscript{o}, the accuracy decreases significantly, resulting in increased background noise when such positions are backprojected with the same intensity. To remedy this, \citeauthor{yoshida2020whole} introduced a \ac{DRF} model specifically designed to incorporate blurring effects along the \ac{LOR} that arise from energy resolution discrepancies. Our \ac{DREP} module extends these concepts, propagating energy resolution (modeled as a Gaussian distribution with 9\% \ac{FWHM} for 511-keV photons in \ac{XEMIS}) and spatial uncertainties (uniform distribution within 3.125\texttimes{}3.125\texttimes{}0.1~mm\textsuperscript{3} voxels) to estimate the blur corresponding to the uncertainty on the Compton angle. We then construct the histoimage by backprojecting the estimated asymmetric Gaussians. 

For stage (iii), we employ a \ac{3D}  model capable of mapping histoimages to real emission sites in real time.

This paper is structured as follows. Section~\ref{sec:method} details the complete pipeline from raw data to reconstructed images. Section~\ref{sec:experiments} describes our experimental setup to assess the accuracy of \ac{MIN} for photon interaction sequence determination as well as the accuracy of Direct3$\upgamma$ for image reconstruction. Section~\ref{sec:results} presents the results of our experiments. Section~\ref{sec:discussion} discusses limitations, potential improvements, and alternative approaches. Finally, Section~\ref{sec:conclusion} concludes this paper.

%% file: content/method.tex
\section{Method}\label{sec:method}

The objective is to reconstruct an  \ac{3D} activity image $\boldx \in \R^P$ where $P$ is the number of voxels in the \ac{FOV}, from a collection of $K$ \ac{3g} detection events.

\subsection{Photon Interaction Sequence Determination}\label{sec:sequence}

In this section, we describe our approach to determine the photon interaction sequence in the detector, which is then used to draw the Compton cone. Although the standard approach consists of considering events with only two interactions (which represent $\approx$30\% of events based on our simulations) and assuming that the interaction with the largest energy deposit is the first (which is not guaranteed), our approach is designed to process events with more than two interactions.

A prompt gamma detection event is represented by a collection of $N$ events (interactions) of unknown order, $\boldo_1 = (\boldr_1,E_1),\dots,\boldo_N =(\boldr_N,E_N)$ where for all $k=1,\dots,N$, $\boldr_k = (x_k,y_k,z_k) \in \R^3$ is the \ac{3D} location of the $k$th interaction and $E_k$, $k\ge 1$ is the deposited energy of the photon at the $k$th interaction. In addition to the prompt gamma, two back-to-back 511-keV gamma rays are emitted and are detected at $\boldb_1$ and $\boldb_2$ (see Figure~\ref{Cone_LOR}).

Considering that a prompt gamma interacts $N$ times in the detector, there are $N!$ possible paths. We first describe the $\mathrm{d}\phi$-criterion approach (Section~\ref{sec:dphi}) and a conventional \ac{NN} approach from the literature (Section~\ref{sec:nn}). We then introduce our proposed \ac{MIN} (Section~\ref{sec:GNN_order}).

\subsubsection{The $\rmd \phi$-criterion}\label{sec:dphi}

We denote by $\calE_k$, $k=1,\dots,N-1$, the energy of the photon after the $k$th interaction, and we denote by $\calE_0$ the initial energy of the photon, so that $E_k = \calE_{k-1} - \calE_k$.

In Compton kinematics, the $k$th scattering angle, denoted $\theta_k^{\mathrm{kin}}$, is determined by the Compton scatter formula \cite{compton1923quantum}:
\begin{equation}\label{eq:kn}
	\cos\left(\theta^{\mathrm{kin}}_k \right) = 1 - \frac{m_{\mathrm{e}}c^2 (\calE_{k-1} - \calE_k)  }{\calE_{k-1}\calE_{k} }  ,\quad k=1,\dots, N-1 \, , 
\end{equation}
where $m_\rme$ is the mass of an electron and $c$ is the speed of light.

The $\mathrm{d}\phi$-criterion \cite{oberlack2000compton} evaluates the fit between the geometric angles $\theta^{\mathrm{geom}}_k$ (determined by a given sequence of interactions) and $\theta^{\mathrm{kin}}_k$, $k=2,\dots,N-1$, as
\begin{equation}\label{eq:dphi}
	\mathrm{d}\phi = \sum_{k=2}^{N-1} \left( \cos\left( \theta^{\mathrm{kin}}_k \right) - \cos\left(\theta^{\mathrm{geom}}_k\right) \right)^2 \, .
\end{equation}
The photon interaction sequence is determined by minimizing \eqref{eq:dphi}, which is achieved by computing all possible sequences. Note that the solution is not necessarily unique. Furthermore, poor energy resolution leads to a decrease in the accuracy of the solution. The first scattering angle, $\theta_\rmc = \theta^{\mathrm{kin}}_1 $, is the Compton cone angle and is not part of the $\mathrm{d}\phi$-criterion as the emission site is outside of the detector.

\subsubsection{Fully-Connected Neural Network}\label{sec:nn}

\citeauthor{NNComptonKinematicsar}~\cite{NNComptonKinematicsar} proposed an architecture that takes as input the normalized deposited energy and positions of the interactions combined with other statistics obtained via simulation, such as the Compton scatter angle, the measured total energy, the $\mathrm{d}\phi$-criterion, the distance between the interactions, as well as the absorption and scatter probabilities and the number of interactions. The \ac{FCNN} contains one hidden layer whose task is to classify the right sequence, so that the output layer contains $N!$ neurons, each of which referring to a possible path with a given probability; the path with the highest probability is then selected. \citeauthor{NNComptonKinematicsar} showed that this type of network is suitable for events with $N=3$ or $4$ interactions but may diverge with for $N>4$ due to the complexity of the problem.   

We implemented a modified version of the architecture proposed by \citeauthor{NNComptonKinematicsar} (Figure~\ref{fig:Meth_FCNN_seq}) with deeper layers. Our model is trained on the deposited energy $E_k$ and position coordinates $\boldr_k = (x_k,y_k,z_k)$ only, without providing additional information about other statistics. This simplified architecture is easier to train and will be used for comparison (Section~\ref{sec:results1}).

\begin{figure}[!h]
	\centering

		\input{./tikz/fcnn}
		\label{subfig:Meth_FCNN_seq_Our}
	
	\caption{
		\Ac{FCNN} architectures for photon path estimation:  where $x_i, y_i, z_i$ are the normalized coordinates of each hit, $E_i$ the deposited energy, 
	}
\label{fig:Meth_FCNN_seq}
\end{figure}

\subsubsection{Modified Interaction Network for Sequence Reconstruction (Proposed Method)}\label{sec:GNN_order}

In their original paper, \citeauthor{Interaction_networks}~\cite{Interaction_networks} proposed an architecture, namely \acf{IN}, which takes a directed graph as input and outputs values associated to the nodes, edges, or the entire graph (e.g., graph classification). The method proved to be a powerful general framework for modeling objects and the relations between them.

\citeauthor{andersson2021gamma}~\cite{andersson2021gamma} proposed a \ac{GNN} framework for gamma-ray track reconstruction for germanium detector arrays. In their work, the location of the first interaction is assumed to be known, and the network is tasked with determining the positions and energies of gamma interactions within the detector. The network must also disentangle and build separate tracks for different gamma photons that are emitted simultaneously, ensuring accurate reconstruction even in complex scenarios with multiple interactions. However, in our case, the location of the first interaction is unknown. 

We propose a simple \ac{GNN} architecture using the framework proposed by \citeauthor{Interaction_networks}~\cite{Interaction_networks}. We consider a collection of $N$ nodes $\boldo_1,\dots,\boldo_N$, with $N_\rmf$ `features', that is, for all $k=1,\dots,N$, $\boldo_k \in \R^{N_\rmf}$, and $N_\rme = N(N-1)$ possible edges (see Figure~\ref{subfig:I_Graph} for the case $N=3$. The nodes are concatenated into a matrix $\boldO = [\boldo_1\transp , \dots ,\boldo_N\transp ]\transp \in \R^{N\times N_\rmf}$ where `$\transp$' denotes the matrix transposition operation.  In our case, $\boldo_k = (\boldr_k, E_k)$ (the \ac{3D} coordinates of the interaction and the deposited energy) and the number of features per node is $N_\rmf = 4$. We utilize a method based on \acp{IN} to determine photon interaction sequences from $\boldO$. This approach treats the problem as an edge classification task in a graph-structured representation of the interactions so that the model can learn complex patterns of energy deposits and scattering angles without explicitly encoding the laws of physics \cite{andersson2021gamma}. Furthermore, \acp{GNN} can process a variable number of interaction points per event and scales well to large numbers of interaction points better than \ac{FCNN}. Note that for the purpose of Compton cone determination, only the first two interactions need to be determined.

The overall procedure, summarized in Figure~\ref{Classification_architecture} and fully described in Appendix~\ref{sec:classification}, defines a mapping $\boldF_{\boldvartheta} \colon \R^{N\times N_\rmf}  \to [0,1]^{N_\rme}$ that maps a collection of interactions $\boldo_1=(\boldr_1,E_1),\dots,\boldo_N = (\boldr_N,E_N)$  to a $N_\rme$-dimensional fuzzy vector containing  probabilities for each edge. Edges with a  probability lower than 0.5 are then removed, which returns the final graph (Figure~\ref{subfig:O_Graph} shows a possible output).

Note that $\boldF_{\boldvartheta}$ does not discard non-admissible graphs, that is to say, containing V-structures (two edges departing from one node or two edges pointing to one node) and cycles. However, as we will see in Section~\ref{sec:results1}, a well-trained network is unlikely to return such graphs. In addition, the architecture of $\boldF_{\boldvartheta}$ does not guarantee that two permutation-equivariant collections of nodes $\boldO$ and $\boldO'$ are mapped to the same graph. We will also show in Section~\ref{sec:results1} that the same graphs (after binarization) are obtained in most cases.

The model is trained from a collection of interaction/graph pairs  $(\boldO , \boldy)$, $\boldy \in \{0,1\}^{N_\rme}$ being a binary vector corresponding to the true sequence of interactions ($[\boldy]_\ell = 1$ if the edge $\boldy_\ell$ is present and $[\boldy]_\ell = 0$ otherwise---Figure~\ref{subfig:O_Graph} shows the graph corresponding to $\boldy = [0,0,1,0,0,1]\transp$),  
obtained from \ac{MC} simulations performed with the \ac{GATE} toolkit, by minimizing the cross-entropy  between the estimated (fuzzy) graph $\boldF_{\boldvartheta} (\boldO)$ and $\boldy$
\begin{align}
	\min_{\boldvartheta}  \, & \mathbb{E}_{(\boldO , \boldy)} [\calL(\boldF_{\boldvartheta} (\boldO),\boldy)] \, ,  \nonumber \\
    \mathcal{L}(\boldz, \boldy) \triangleq & - \sum_{l=1}^{N_{\rme}} y_l \log z_l + (1-y_l) \log(1-z_l) \, \forall \boldy,\boldz\in\R^{N_{\rme}} \, .   \nonumber
\end{align}

\begin{figure*}
	\includegraphics[width=\textwidth]{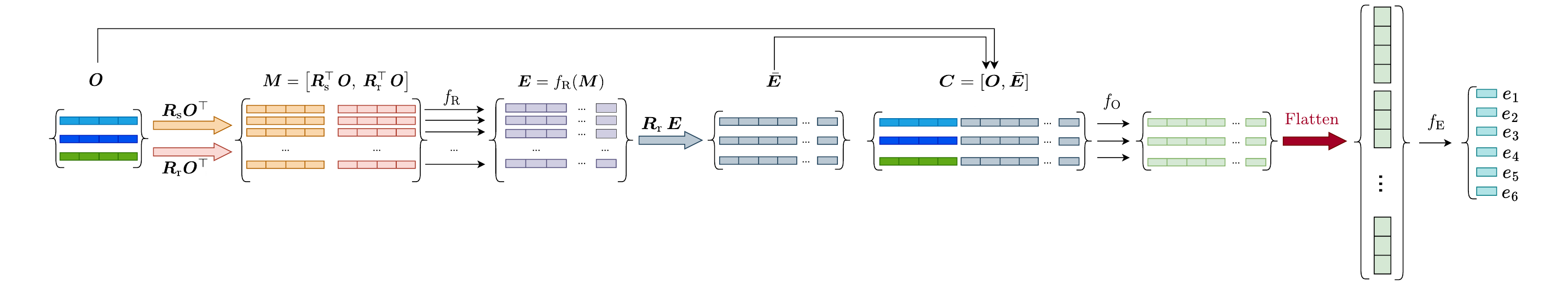}
	\caption{\Ac{MIN}---Proposed architecture for $\boldF_{\boldvartheta}$ used to classify edges or relations between photons interactions in the detector (see Appendix~\ref{sec:classification}).}
	\label{Classification_architecture}
\end{figure*}

\subsection{Histoimage Generation}\label{sec:emission}

\subsubsection{Emission Point Estimation}

Using the first two interaction locations $\boldr_1  $ and $\boldr_2$ (which were determined using \ac{MIN}, cf. Section~\ref{sec:GNN_order}) and the measured deposited energy $E_1$ in $\boldr_1$, the Compton cone angle $\theta_\rmc = \theta^{\mathrm{kin}}_1$ can be determined using the Compton scatter formula \eqref{eq:kn} with $\calE_0=1.157 \, \text{MeV}$ and $\calE_1= \calE_0 - E_1$.
%
%
The cone vertex is located at the first interaction point $\boldr_1$, and its axis runs through to the second interaction point $\boldr_2$ (Figure~\ref{Cone_LOR}).

The emission point of a prompt gamma  can be estimated as the intersection point between the Compton cone and the \ac{LOR} given by the two points $\boldb_1$ and $\boldb_2$, i.e., by solving
\begin{equation}
	\text{find $\boldp \in (\boldb_1,\boldb_2)$ s.t.} \quad \frac{(\boldp - \boldr_1) \cdot \boldn}{\|\boldp - \boldr_1\|_2} = \cos(\theta_c) 
	\label{eq:Cone_solution}
\end{equation}
where $\boldn = ( \boldr_2-\boldr_1) / \|\boldr_2-\boldr_1  \|_2$. 

\subsubsection{Modeling Spatial Uncertainty}

The utilization of this method is challenging due to the imperfections in the detector. In the case of the \ac{XEMIS} detector, significant uncertainties exist in both energy and spatial measurements. The detector exhibits an energy resolution of 9\% \ac{FWHM} for 511-keV $\upgamma$-rays, modeled as a Gaussian distribution around measured energy values.  The intrinsic spatial resolution is determined by the detector size and is assumed to follow a uniform distribution.

Translating the uncertainties associated with the detector measurements into the cone angle $\theta_\rmc$ results in uncertainty around the estimated point $\boldp$ on the \ac{LOR}. However, using a symmetric error model around the estimated point, as described in \citeauthor{Giovagnoli}~\cite{Giovagnoli}, is not the most effective method to transfer this uncertainty from the Compton cone to the \ac{LOR}. The uncertainty along the \ac{LOR} is influenced by several factors, including the crossing angle between the Compton cone and the \ac{LOR} as well as the distance between the vertex of the cone and the \ac{LOR}.

Based on the findings from \citeauthor{yoshida2020whole}~\cite{yoshida2020whole}, we propose to use a nonsymmetric `Gaussian' kernel $h_{\boldp}$ around the estimated point of emission $\boldp$ :  
\begin{equation}
	h_{\boldp}(t) = 
	\begin{cases} 
		\frac{1}{\sqrt{2\pi} (\sigma_+ + \sigma_-)} \exp\left(-\frac{(t - t_{\boldp})^2}{2\sigma_-^2}\right) & \text{for } t < t_{\boldp}, \\
		\frac{1}{\sqrt{2\pi} (\sigma_+ + \sigma_-)} \exp\left(-\frac{(t - t_{\boldp})^2}{2\sigma_+^2}\right) & \text{for } t \geq t_{\boldp}.
	\end{cases}
	\label{eq:sigma_uniform}
\end{equation}
where $t_{\boldp}\in\R$ is the position of $\boldp$ along the \ac{LOR} and $\sigma_+$ and $\sigma_-$ are calculated by \ac{DREP}  (cf. Appendix~\ref{sec:uncertainty}, Equation~\eqref{eq:sigma+-}).

\subsubsection{Attenuation Correction}\label{sec:attenuation_correction}

In \ac{3g} \ac{PET} imaging with \textsuperscript{44}Sc, it is necessary to correct for attenuation of 511-keV and 1.157-MeV gamma rays. The attenuation correction factors $a_{\boldp}$ for each $\boldp$ is given by
\begin{equation}\label{eq:attn_factor}
	a_{\boldp} = \rme^{\int_{\bm{b}_1}^{\bm{b}_2} \mu_{511}(\boldr) \, \rmd{}\boldr} \cdot \rme^{\int_{\boldp}^{\boldr_1} \mu_{1157}(\boldr) \, \rmd{}\boldr}  \notag
\end{equation}
where $\mu_{511}(\boldr)$ and $\mu_{1157}(\boldr)$ are the attenuation coefficients at position $\boldr\in\R^3$ for 511-keV and 1.157-MeV gamma rays, respectively. The attenuation-corrected kernel  $h_{\boldp}^\mathrm{ac}$ is defined as
\begin{equation}
	h_{\boldp}^\mathrm{ac} = a_{\boldp} \cdot h_{\boldp} \, .  \nonumber
\end{equation}

\subsubsection{Three-gamma Histogrammer}\label{sec:histogrammer}

\begin{figure*}	\includegraphics[width=\textwidth]{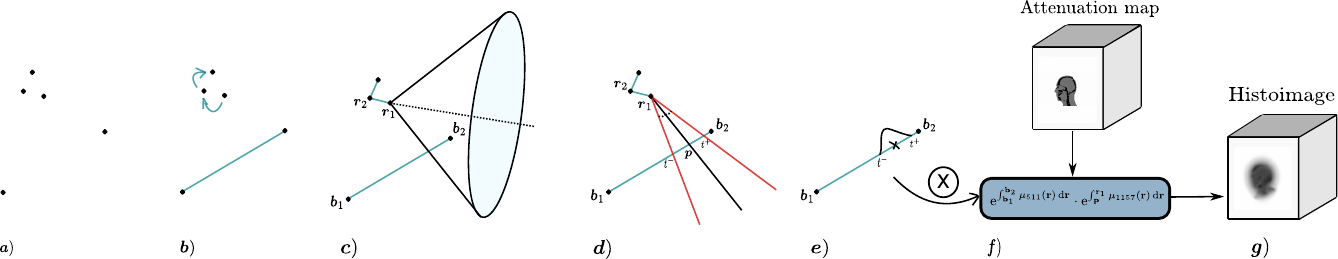}
	\caption{
		Workflow of the Direct3$\upgamma$ histogrammer: 
		(a) detection of hits by the \ac{3g} \ac{PET} scanner; 
		(b) construction of \ac{LOR} and determination of prompt gamma interaction order; 
		(c) estimation of intersection between Compton cone and \ac{LOR}; 
		(d) calculation of uncertainties on the \ac{LOR}; 
		(e) projection of estimated Gaussian distribution onto image space; 
		and (f) application of attenuation correction. 
	}
\label{fig:histogrammer}
\end{figure*}

We define a \ac{3g} histogrammer using a non-symmetric Gaussian \ac{PDF} to model the emission event distribution along the \ac{LOR}. This approach extends the `Most Likely Annihilation Position histogrammer'  proposed by \citeauthor{whiteley2020fastpet}~\cite{whiteley2020fastpet} for \ac{TOF} \ac{PET}. Given a collection of emission point $\{\boldp_1,\dots,\boldp_K\}$ estimated from the $K$ \ac{3g} events, the \ac{3g} histoimage $\boldx^{\mathrm{hist}}\in\R^P$ ($P$ being the number of voxels) is created by summing voxelized $P$-dimensional versions of the attenuation-corrected kernels $h_{\boldp_k}^\mathrm{ac}$, denoted $\boldh_k^\mathrm{ac}\in\R^P$:
\begin{equation}\nonumber
	\boldx^{\mathrm{hist}} = \sum_{k=1}^K \boldh_k^\mathrm{ac}.
\end{equation}
The overall histogramming procedure is summarized in Figure~\ref{fig:histogrammer}.

\subsection{From the Histoimage to the Final Image}\label{sec:finalstep}

This section describe the methodology to derive the final image $\boldx$ from the histoimage  $\boldx^{\mathrm{hist}}$.

Equation~\eqref{eq:Cone_solution} can result in two possible intersection points between the Compton cone and the \ac{LOR} within the \ac{FOV}. Both of these solutions are utilized in the creation of histoimages, leading to noise caused by false positives in $\boldx^{\mathrm{hist}}$. Another possible source of noise are the inaccuracies in the algorithm responsible for determining the sequence of detected gamma rays. Furthermore, we need to correct for the uncertainty along the \ac{LOR} in the histoimages.

To address the issue of blurring and noise in histoimages, we propose the use of a simple U-Net architecture as the backbone for our image-to-image translation tasks (the complete pipeline is illustrated in Figure~\ref{fig:DIRECT3gammaFinal}). In this work, we implemented a generative model,  which combines the U-Net architecture with a patch discriminator and a least-squares \ac{GAN} loss function, following the approach of \citeauthor{isola2017image}~\cite{isola2017image} and \citeauthor{cirillo2021vox2vox}~\cite{cirillo2021vox2vox}. In this generative framework, the U-Net $\boldG_{\boldpsi} \colon \R^P \to \R^P$, with trainable parameter $\boldpsi$, maps the histoimage $\boldx^{\mathrm{hist}} \in \R^P$ to the true emission image $\boldx^\star \in \R^P$. The model learns to refine the image by reducing noise and blur, while the patch-based discriminator with adversarial training ensures that the output images hold high-frequency details.

%% file: tikz/fcnn.tex
	
\begin{tikzpicture}[scale=0.6, transform shape]
	\small
	\newlength{\dimneuron}
	\setlength{\dimneuron}{10pt}

	\newlength{\shiftneuron}
	\setlength{\shiftneuron}{5pt}
	
	\newlength{\spaceneuron}
	\setlength{\spaceneuron}{25pt}
	
	\newlength{\dimdot}
	\setlength{\dimdot}{2pt}

	\newlength{\spacelabelin}
	\setlength{\spacelabelin}{38pt}
	
	\newlength{\spacelabelout}
	\setlength{\spacelabelout}{32pt}

	\newlength{\spacelayer}
	\setlength{\spacelayer}{60pt}
	
	\newlength{\connector}
	\setlength{\connector}{0.2pt}
	
	
	\node(x1) at (0,0) [circle,inner sep=0pt,minimum size=\dimneuron,draw,thick,c2,fill=c22] {} ;
	\node(x2) at ([yshift=-\spaceneuron]x1) [circle,inner sep=0pt,minimum size=\dimneuron,draw,thick,c2,fill=c22] {} ;
	\node(x3) at ([yshift=-\spaceneuron]x2) [circle,inner sep=0pt,minimum size=\dimneuron,draw,thick,c2,fill=c22] {} ;

	\node(dot1) at ([yshift=-\spaceneuron]x3) [circle,inner sep=0pt,minimum size=\dimdot,draw,fill] {} ;
	\node(dot2) at ([yshift=-\spaceneuron/2]dot1) [circle,inner sep=0pt,minimum size=\dimdot,draw,fill] {} ;
	\node(dot3) at ([yshift=-\spaceneuron/2]dot2) [circle,inner sep=0pt,minimum size=\dimdot,draw,fill] {} ;
	
	\node(xN) at ([yshift=-\spaceneuron]dot3) [circle,inner sep=0pt,minimum size=\dimneuron,draw,thick,c2,fill=c22] {} ;
	
	\node(labelx1) at ([xshift=-\spacelabelin]x1) [] {$x_1,y_1,z_1,E_1$} ;
	\node(labelx2) at ([xshift=-\spacelabelin]x2) [] {$x_2,y_2,z_2,E_2$} ;
	\node(labelx3) at ([xshift=-\spacelabelin]x3) [] {$x_3,y_3,z_3,E_3$} ;
	\node(labelN) at ([xshift=-\spacelabelin]xN) [] {$x_N,y_N,z_N,E_N$} ;
	
	
	\node(y1) at ([xshift=\spacelayer,yshift=-\spaceneuron/2]x1) [circle,inner sep=0pt,minimum size=\dimneuron,draw,thick,c1,fill=c11] {} ;
	\node(y2) at ([yshift=-\spaceneuron]y1) [circle,inner sep=0pt,minimum size=\dimneuron,draw,thick,c1,fill=c11] {} ;
	\node(y3) at ([yshift=-\spaceneuron]y2) [circle,inner sep=0pt,minimum size=\dimneuron,draw,thick,c1,fill=c11] {} ;
	\node(y4) at ([yshift=-\spaceneuron]y3) [circle,inner sep=0pt,minimum size=\dimneuron,draw,thick,c1,fill=c11] {} ;
	\node(y5) at ([yshift=-\spaceneuron]y4) [circle,inner sep=0pt,minimum size=\dimneuron,draw,thick,c1,fill=c11] {} ;
	
	
	\node(z1) at ([xshift=\spacelayer]y1) [circle,inner sep=0pt,minimum size=\dimneuron,draw,thick,c1,fill=c11] {} ;
	\node(z2) at ([yshift=-\spaceneuron]z1) [circle,inner sep=0pt,minimum size=\dimneuron,draw,thick,c1,fill=c11] {} ;
	\node(z3) at ([yshift=-\spaceneuron]z2) [circle,inner sep=0pt,minimum size=\dimneuron,draw,thick,c1,fill=c11] {} ;
	\node(z4) at ([yshift=-\spaceneuron]z3) [circle,inner sep=0pt,minimum size=\dimneuron,draw,thick,c1,fill=c11] {} ;
	\node(z5) at ([yshift=-\spaceneuron]z4) [circle,inner sep=0pt,minimum size=\dimneuron,draw,thick,c1,fill=c11] {} ;
	
	
	\node(t1) at ([xshift=\spacelayer]z1) [circle,inner sep=0pt,minimum size=\dimneuron,draw,thick,c1,fill=c11] {} ;
	\node(t2) at ([yshift=-\spaceneuron]t1) [circle,inner sep=0pt,minimum size=\dimneuron,draw,thick,c1,fill=c11] {} ;
	\node(t3) at ([yshift=-\spaceneuron]t2) [circle,inner sep=0pt,minimum size=\dimneuron,draw,thick,c1,fill=c11] {} ;
	\node(t4) at ([yshift=-\spaceneuron]t3) [circle,inner sep=0pt,minimum size=\dimneuron,draw,thick,c1,fill=c11] {} ;
	\node(t5) at ([yshift=-\spaceneuron]t4) [circle,inner sep=0pt,minimum size=\dimneuron,draw,thick,c1,fill=c11] {} ;
	
	
	\node(o1) at ([xshift=\spacelayer,yshift=\spaceneuron/2]t1) [circle,inner sep=0pt,minimum size=\dimneuron,draw,thick,c3,fill = c33] {} ;
	\node(o2) at ([yshift=-\spaceneuron]o1) [circle,inner sep=0pt,minimum size=\dimneuron,draw,thick,c3,fill = c33] {} ;
	\node(o3) at ([yshift=-\spaceneuron]o2) [circle,inner sep=0pt,minimum size=\dimneuron,draw,thick,c3,fill = c33] {} ;
	
	\node(dot1) at ([yshift=-\spaceneuron]o3) [circle,inner sep=0pt,minimum size=\dimdot,draw,fill] {} ;
	\node(dot2) at ([yshift=-\spaceneuron/2]dot1) [circle,inner sep=0pt,minimum size=\dimdot,draw,fill] {} ;
	\node(dot3) at ([yshift=-\spaceneuron/2]dot2) [circle,inner sep=0pt,minimum size=\dimdot,draw,fill] {} ;
	
	\node(oN) at ([yshift=-\spaceneuron]dot3) [circle,inner sep=0pt,minimum size=\dimneuron,draw,thick,c3,fill = c33] {} ;
	
	\node(labelo1) at ([xshift=\spacelabelout]o1) [] {Sequence 1} ;
	\node(labelo2) at ([xshift=\spacelabelout]o2) [] {Sequence 2} ;
	\node(labelo3) at ([xshift=\spacelabelout]o3) [] {Sequence 3} ;
	\node(labeloN) at ([xshift=\spacelabelout]oN) [] {Sequence $N!$} ;
	
	
	\draw[->,line width=\connector]   (x1)--(y1) ;
	\draw[->,line width=\connector]   (x1)--(y2) ;
	\draw[->,line width=\connector]   (x1)--(y3) ;
	\draw[->,line width=\connector]   (x1)--(y4) ;
	\draw[->,line width=\connector]   (x1)--(y5) ;
	
	\draw[->,line width=\connector]   (x2)--(y1);
	\draw[->,line width=\connector]   (x2)--(y2);
	\draw[->,line width=\connector]   (x2)--(y3);
	\draw[->,line width=\connector]   (x2)--(y4);
	\draw[->,line width=\connector]   (x2)--(y5) ;
	
	\draw[->,line width=\connector]   (x3)--(y1) ;
	\draw[->,line width=\connector]   (x3)--(y2) ;
	\draw[->,line width=\connector]   (x3)--(y3) ;
	\draw[->,line width=\connector]   (x3)--(y4) ;
	\draw[->,line width=\connector]   (x3)--(y5) ;
	
	\draw[->,line width=\connector]   (xN)--(y1) ;
	\draw[->,line width=\connector]   (xN)--(y2) ;
	\draw[->,line width=\connector]   (xN)--(y3) ;
	\draw[->,line width=\connector]   (xN)--(y4) ;
	\draw[->,line width=\connector]   (xN)--(y5) ;
	
	
	\draw[->,line width=\connector]   (y1)--(z1) ;
	\draw[->,line width=\connector]   (y1)--(z2) ;
	\draw[->,line width=\connector]   (y1)--(z3) ;
	\draw[->,line width=\connector]   (y1)--(z4) ;
	\draw[->,line width=\connector]   (y1)--(z5) ;
	
	\draw[->,line width=\connector]   (y2)--(z1) ;
	\draw[->,line width=\connector]   (y2)--(z2) ;
	\draw[->,line width=\connector]   (y2)--(z3) ;
	\draw[->,line width=\connector]   (y2)--(z4) ;
	\draw[->,line width=\connector]   (y2)--(z5) ;
	
	\draw[->,line width=\connector]   (y3)--(z1) ;
	\draw[->,line width=\connector]   (y3)--(z2) ;
	\draw[->,line width=\connector]   (y3)--(z3) ;
	\draw[->,line width=\connector]   (y3)--(z4) ;
	\draw[->,line width=\connector]   (y3)--(z5) ;
	
	\draw[->,line width=\connector]   (y4)--(z1) ;
	\draw[->,line width=\connector]   (y4)--(z2) ;
	\draw[->,line width=\connector]   (y4)--(z3) ;
	\draw[->,line width=\connector]   (y4)--(z4) ;
	\draw[->,line width=\connector]   (y4)--(z5) ;
	
	\draw[->,line width=\connector]   (y5)--(z1) ;
	\draw[->,line width=\connector]   (y5)--(z2) ;
	\draw[->,line width=\connector]   (y5)--(z3) ;
	\draw[->,line width=\connector]   (y5)--(z4) ;
	\draw[->,line width=\connector]   (y5)--(z5) ;
	
	
	\draw[->,line width=\connector]   (z1)--(t1) ;
	\draw[->,line width=\connector]   (z1)--(t2) ;
	\draw[->,line width=\connector]   (z1)--(t3) ;
	\draw[->,line width=\connector]   (z1)--(t4) ;
	\draw[->,line width=\connector]   (z1)--(t5) ;
	
	\draw[->,line width=\connector]   (z2)--(t1) ;
	\draw[->,line width=\connector]   (z2)--(t2) ;
	\draw[->,line width=\connector]   (z2)--(t3) ;
	\draw[->,line width=\connector]   (z2)--(t4) ;
	\draw[->,line width=\connector]   (z2)--(t5) ;
	
	\draw[->,line width=\connector]   (z3)--(t1) ;
	\draw[->,line width=\connector]   (z3)--(t2) ;
	\draw[->,line width=\connector]   (z3)--(t3) ;
	\draw[->,line width=\connector]   (z3)--(t4) ;
	\draw[->,line width=\connector]   (z3)--(t5) ;
	
	\draw[->,line width=\connector]   (z4)--(t1) ;
	\draw[->,line width=\connector]   (z4)--(t2) ;
	\draw[->,line width=\connector]   (z4)--(t3) ;
	\draw[->,line width=\connector]   (z4)--(t4) ;
	\draw[->,line width=\connector]   (z4)--(t5) ;
	
	\draw[->,line width=\connector]   (z5)--(t1) ;
	\draw[->,line width=\connector]   (z5)--(t2) ;
	\draw[->,line width=\connector]   (z5)--(t3) ;
	\draw[->,line width=\connector]   (z5)--(t4) ;
	\draw[->,line width=\connector]   (z5)--(t5) ;
	
	
	\draw[->,line width=\connector]   (t1)--(o1) ;
	\draw[->,line width=\connector]   (t1)--(o2) ;
	\draw[->,line width=\connector]   (t1)--(o3) ;
	\draw[->,line width=\connector]   (t1)--(oN) ;

	\draw[->,line width=\connector]   (t2)--(o1) ;
	\draw[->,line width=\connector]   (t2)--(o2) ;
	\draw[->,line width=\connector]   (t2)--(o3) ;
	\draw[->,line width=\connector]   (t2)--(oN) ;

	\draw[->,line width=\connector]   (t3)--(o1) ;
	\draw[->,line width=\connector]   (t3)--(o2) ;
	\draw[->,line width=\connector]   (t3)--(o3) ;
	\draw[->,line width=\connector]   (t3)--(oN) ;

	\draw[->,line width=\connector]   (t4)--(o1) ;
	\draw[->,line width=\connector]   (t4)--(o2) ;
	\draw[->,line width=\connector]   (t4)--(o3) ;
	\draw[->,line width=\connector]   (t4)--(oN) ;

	\draw[->,line width=\connector]   (t5)--(o1) ;
	\draw[->,line width=\connector]   (t5)--(o2) ;
	\draw[->,line width=\connector]   (t5)--(o3) ;
	\draw[->,line width=\connector]   (t5)--(oN) ;

\end{tikzpicture}
	

%% file: content/experiments.tex
\section{Experimental Setup}\label{sec:experiments}

\subsection{Detector}

Data acquisition was carried out using simulations implemented through \ac{GATE}  version 9.1 \cite{jan2004gate}, employing the \verb|emstandard_opt3| physics list. focusing on two types of scanner: (i) a human-sized scanner (described below) and (ii) a small-animal (mouse-sized) pre-clinical scanner \ac{XEMIS} \cite{aprile2010liquid, zhu2018scintillation, manzano2016optimization}. 

The human-sized scanner, inspired by the \ac{XEMIS} series, features an \ac{LXe}-based detection system with an inner diameter of 60~cm and an outer diameter of 90~cm. It achieves accurate \ac{3D} localization of photon interactions using a \ac{LXe}-\ac{TPC}. The $(x,y)$ coordinates are determined by tracking the ionization electrons as they drift toward a segmented anode plane with 3.125\texttimes{}3.125~mm\textsuperscript{2} pixels, while the $z$ coordinate is obtained from the time difference between the prompt scintillation signal and electron arrival. This system ensures a longitudinal spatial resolution below 100~\textmu{}m and an energy resolution of ~9\% \ac{FWHM} at 511~keV, which we used to define the energy error $\Delta\calE_1$ in Equation~\eqref{eq:delta} (Appendix~\ref{sec:uncertainty}). To model spatial resolution within the detector, we assume a uniform voxel grid of 3.125\texttimes{}3.125\texttimes{}0.1~mm\textsuperscript{3}, where each interaction is uniformly distributed within its voxel. 

We also designed human- and mouse-sized \ac{TOF} scanners for comparison. The human-sized scanner was designed as a Biograph mMR PET/MRI scanner (Siemens Healthineers), while the mouse-sized scanner is derived from the \ac{XEMIS}  but equipped with \ac{LSO} crystals os size of 1\texttimes{}1\texttimes{}10~mm\textsuperscript{3}. Both scanners have a 200-ps \ac{TOF} resolution. Reconstruction data were simulated from the same phantoms and with the same number of events as \ac{3g} acquisitions.

\subsection{Experiment 1---Photon Interaction Sequence Determination}\label{par:experiment1}

To train the \ac{MIN} and \ac{FCNN} models, a simulation was performed using a uniform cylinder filled with water and \textsuperscript{44}Sc, which occupies the entire \ac{FOV} of each scanner. This simulation generated up to 20 million \ac{3g} events, with only those events included where all gamma rays fully interacted with the detector, ensuring that the total deposited energy across all interactions was 1.157 MeV. Separate models of \ac{MIN} and \ac{FCNN} were created, each tailored to the number of interactions per event. For testing, a distinct dataset of two million events was used, also derived from a uniform cylinder within the \ac{FOV}.

\subsection{Experiment 2---Image Reconstruction}\label{par:experiment23}

A set of 400 anthropomorphic phantoms, generated with the \ac{XCAT} software \cite{segars20104d} with different morphologies, was used to simulate \textsuperscript{44}Sc activity along-side their 511-keV and 1.157-MeV attenuation maps. These phantoms were designed with dimensions of 200\texttimes{}200\texttimes{}200 voxels, with a voxel size of 3\texttimes{}3\texttimes{}3~mm\textsuperscript{3} for the human-sized and 0.8\texttimes{}0.8\texttimes{}0.8~mm\textsuperscript{3} for the small scanner (to produce mouse-sized \ac{XCAT} phantoms), allowing the representation of a wide range of anatomical sizes, shapes, and regions. To simulate various clinical scenarios, spherical lesions of various sizes and shapes were randomly introduced into different regions of the phantoms. For each phantom, between five and $K=20\cdot10^6$ \ac{3g} events (excluding randoms and scatter coincidences) were simulated, and the corresponding histoimages were produced using our histogrammer described in Section~\ref{sec:histogrammer}. 

\begin{figure}[h!]
	
	\centering
	\begin{subfigure}[b]{0.09\textwidth}
		\reflectbox{\includegraphics[width=\textwidth, trim={30 30 30 30}, angle=180, clip]{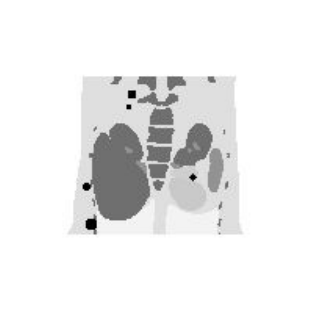}}
		\caption{\scriptsize Phantom 1}
		\label{subfig:Phantom1}
		
	\end{subfigure}
	\begin{subfigure}[b]{0.09\textwidth}
		\reflectbox{\includegraphics[width=\textwidth, trim={30 30 30 30}, angle=180, clip]{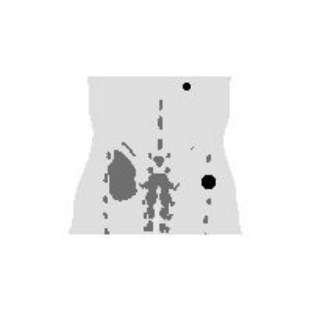}}
		\caption{\scriptsize Phantom 2}
		\label{subfig:Phantom2}
	\end{subfigure}
	\begin{subfigure}[b]{0.09\textwidth}
		\reflectbox{\includegraphics[width=\textwidth, trim={30 30 30 30}, angle=180, clip]{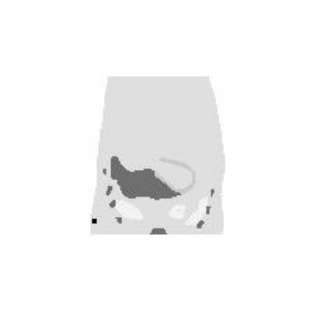}}
		\caption{\scriptsize Phantom 3}\label{subfig:Phantom3}
	\end{subfigure}
	\begin{subfigure}[b]{0.09\textwidth}
		\reflectbox{\includegraphics[width=\textwidth, trim={30 30 30 30}, angle=180, clip]{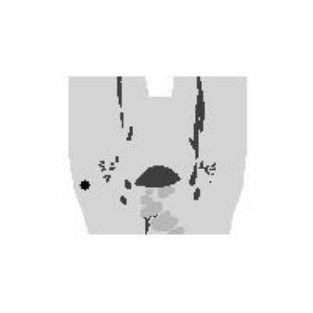}}
		\caption{\scriptsize Phantom 4}\label{subfig:Phantom4}
	\end{subfigure}
	\begin{subfigure}[b]{0.09\textwidth}
		\reflectbox{\includegraphics[width=\textwidth, trim={30 30 30 30}, angle=180, clip]{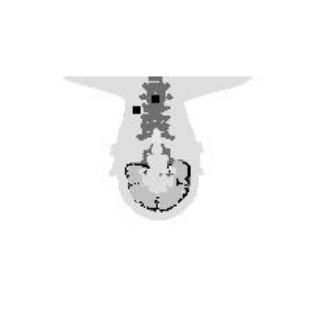}}
		\caption{\scriptsize Phantom 5}\label{subfig:Phantom5}
	\end{subfigure}
	
	
	\begin{subfigure}[b]{0.09\textwidth}
		\reflectbox{\includegraphics[width=\textwidth, trim={30 30 30 30}, angle=180, clip]{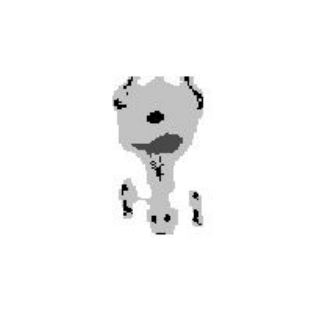}}
		\caption{\scriptsize Phantom 1}
		\label{subfig:Phantom1m}
	\end{subfigure}
	\begin{subfigure}[b]{0.09\textwidth}
		\reflectbox{\includegraphics[width=\textwidth, trim={30 30 30 30}, angle=180, clip]{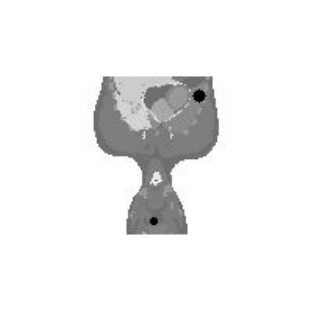}}
		\caption{\scriptsize Phantom 2}
		\label{subfig:Phantom2m}
	\end{subfigure}
	\begin{subfigure}[b]{0.09\textwidth}
		\reflectbox{\includegraphics[width=\textwidth, trim={30 30 30 30}, angle=180, clip]{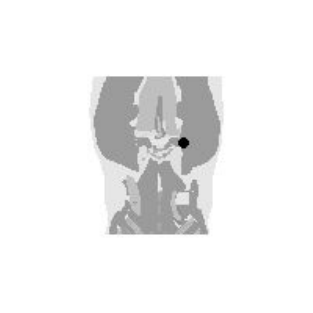}}
		\caption{\scriptsize Phantom 3}\label{subfig:Phantom3m}
	\end{subfigure}
	\begin{subfigure}[b]{0.09\textwidth}
		\reflectbox{\includegraphics[width=\textwidth, trim={30 30 30 30}, angle=180, clip]{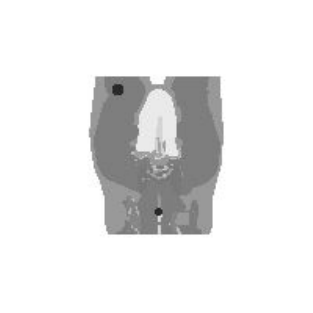}}
		\caption{\scriptsize Phantom 4}\label{subfig:Phantom4m}
	\end{subfigure}
	\begin{subfigure}[b]{0.09\textwidth}
		\reflectbox{\includegraphics[width=\textwidth, trim={30 30 30 30}, angle=180, clip]{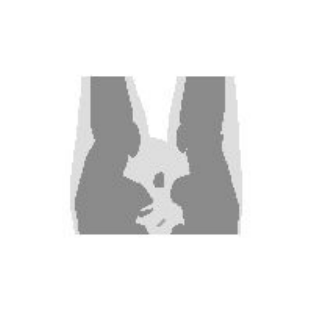}}
		\caption{\scriptsize Phantom 5}\label{subfig:Phantom5m}
	\end{subfigure}
	
	\caption{Coronal views of \subref{subfig:Phantom1}--\subref{subfig:Phantom5} the human-sized phantoms  (3\texttimes{}3\texttimes{}3~mm\textsuperscript{3} voxel size) and \subref{subfig:Phantom1m}--\subref{subfig:Phantom5m} the mouse-sized phantoms (0.8\texttimes{}0.8\texttimes{}0.8~mm\textsuperscript{3} voxel size) used in the test dataset.}
	\label{fig:transverse_phantoms_images}
\end{figure}

The \ac{3D} U-Net described in Section~\ref{sec:finalstep} was then trained to map the histoimages to the real emission maps. To enhance robustness and prevent overfitting, a data augmentation protocol was performed during the training process, including random rigid transformations such as flipping, rotations, translations, and intensity rescaling within (50\% to 250\%). The evaluation on the human-sized was conducted using five distinct \ac{XCAT} volumes, each representing different body positions and varying levels of activity. For the mouse-sized scanner, we used a mouse phantom, as well as four mouse-sized \ac{XCAT} phantoms (mini humans). The transverse views of these phantoms are presented in Figure~\ref{fig:transverse_phantoms_images}.

The Direct3$\upgamma$ reconstruction method was compared with conventional \ac{MLEM}-reconstructed images (80 iterations) from  \ac{TOF} \ac{PET} data. We also considered a \ac{DIP}-processed version of the \ac{TOF}-reconstructed images, namely \ac{DIP}-\ac{TOF}. Our \ac{DIP}-\ac{TOF} framework used a U-Net $\boldF_{\boldtheta}\colon\R^P\to\R^P$ with an architecture similar to that proposed in \citeauthor{ulyanov2018deep}~\cite{ulyanov2018deep}. Given a random input image $\boldz\in\R^P$ and the initial \ac{MLEM}-\ac{TOF} estimate $\tilde{\boldx}$, the \ac{DIP}-\ac{TOF} reconstructed image $\hat{\boldx}$ is given by
\begin{align}
    \hat{\boldtheta} \in {} & \argmin_{\boldtheta} \, \| \tilde{\boldx} - \boldF_{\boldtheta}(\boldz) \|^2 \label{eq:dip} \\
    \hat{\boldx} = {} & \boldF_{\hat{\boldtheta}}(\boldz) \, .
\end{align}
We implemented \ac{DIP} with a \ac{3D} U-Net, 600 epochs and a learning rate of $10^{-3}$ to solve \eqref{eq:dip} to balance between noise control and structure preservation and to prevent overfitting.

%% file: content/results.tex
\section{Results}\label{sec:results}

\subsection{Experiment 1---Photon Interaction Sequence Determination}\label{sec:results1}

In this section, we show the results on prediction of the order of interaction of the three methods described in Section~\ref{sec:sequence}, i.e.,  $\rmd\phi$-criterion, \ac{FCNN} and \ac{MIN} (proposed approach), from the simulated data (Section~\ref{par:experiment1}).

Table~\ref{tab:accuracy} presents the accuracy of the three algorithms in predicting photon interaction sequences. The table shows results for events with three, four, and five interactions. For events with only two interactions, no algorithm was applied; instead, we simply chose the position with the highest energy as the first interaction, resulting in approximately 81\% accuracy based on our simulation. The table includes two additional columns: ``All Events'' and ``First two only''. The column ``All Events'' represents the overall precision in reconstructing the entire interaction path for all events. The column ``First two only'' shows the precision in reconstructing the first and second points of the sequence only (which are sufficient to draw the Compton cone). The accuracy is calculated as
\begin{equation}
    \textrm{Accuracy} = \frac{\textrm{Number of well classified sequences}}{\textrm{Total number of events}}  \, \notag .
\end{equation}

\begin{table}[h!]
    \centering
    \begin{tabular}{ccccc}
        \toprule
        \textbf{Approach} & $\bm{N=3}$ & $\bm{N=4}$ & $\bm{N=5}$ & 
        \textbf{First 2 only} \\
        \midrule
        $\rmd \phi$-criterion     & 88   & 73.5 & 61   
        & 0.798 \\
        \ac{FCNN}       & 91   & 82   & 59   
        & 0.82 \\
        \ac{MIN}    & \bf 93.5 & \bf 92   & \bf 77   
        & \bf 0.877 \\
        \bottomrule
    \end{tabular}
    \caption{Experiment 1---Comparison of the three different approaches ($\rmd \phi$-criterion, \ac{FCNN}, \ac{MIN}) for predicting the photon interaction sequence.}
    \label{tab:accuracy}
\end{table}

The results in Table~\ref{tab:accuracy} demonstrate that the \ac{MIN} approach consistently outperforms the other two methods in all scenarios. It achieves the highest accuracy for events with three, four, and five interactions,  and when considering only the first two interactions. As the number of interactions increases, the accuracy of all methods decreases, indicating that longer interaction sequences are more challenging to reconstruct. Interestingly, all methods show slightly higher accuracy when focusing on just the first two interactions, which is crucial for Compton cone reconstruction. It is important to note that \ac{MIN} produced nonadmissible sequences (V-structures, cycles) for 2\% of the events, which were included in the accuracy calculations. Despite this limitation, \ac{MIN} remains the most effective approach to predict photon interaction sequences in this study.

\subsection{Experiment 2---Image Reconstruction}\label{sec:results2}

The reconstructed images were assessed against the real emission map using \acs{SSIM} and \acs{PSNR} as quantitative evaluation metrics.

\subsubsection{Human-sized Scanner}

\begin{figure*}[htbp]
	\centering
	\begin{tabular}{c cccccccc}
		
		& \scriptsize \acs{GT} & \scriptsize Direct3$\upgamma$ & \scriptsize Direct3$\upgamma$ Error & \scriptsize \ac{TOF} & \scriptsize \ac{TOF} Error & \scriptsize \ac{DIP}-\ac{TOF} &\scriptsize  \ac{DIP}-\ac{TOF} Error \\
		
		\multirow{2}{*}[+6ex]{\centering \rotatebox{90}{\scriptsize Saggital}} &
		\begin{overpic}[width=0.11\textwidth,angle=180,trim=12 0 12 0,clip]{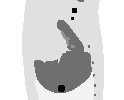}
		\end{overpic} &
		\begin{overpic}[width=0.11\textwidth,angle=180,trim=12 0 12 0,clip]{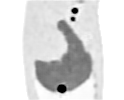}
		\end{overpic} &
		\begin{overpic}[width=0.11\textwidth,angle=180,trim=12 0 12 0,clip]{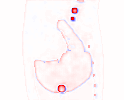}
			\put(80,0){\includegraphics[width=0.03\textwidth,height=0.12\textwidth,trim=0 0 0 0,clip]{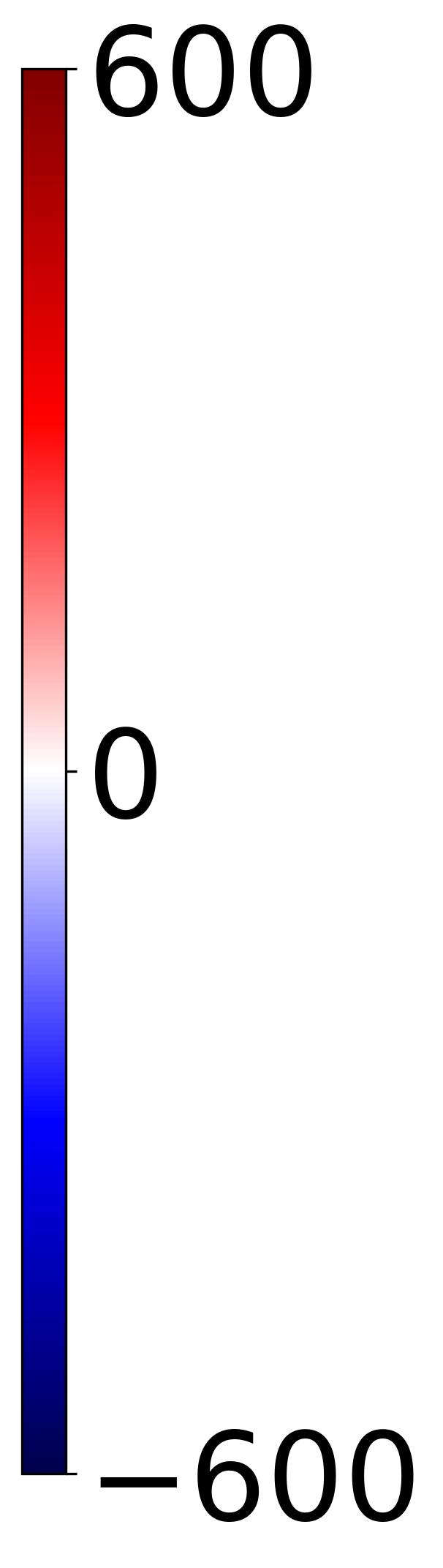}}
		\end{overpic} &
		\begin{overpic}[width=0.11\textwidth,angle=180,trim=12 0 12 0,clip]{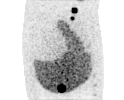}
		\end{overpic} &
		\begin{overpic}[width=0.11\textwidth,angle=180,trim=12 0 12 0,clip]{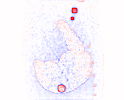}
			\put(80,0){\includegraphics[width=0.03\textwidth,height=0.12\textwidth,trim=0 0 0 0,clip]{images/Image907_2/cbar.png}}
		\end{overpic} &
		\begin{overpic}[width=0.11\textwidth,angle=180,trim=12 0 12 0,clip]{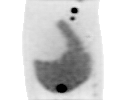}
		\end{overpic} &
		\begin{overpic}[width=0.11\textwidth,angle=180,trim=12 0 12 0,clip]{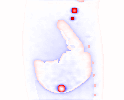}
			\put(90,0){\includegraphics[width=0.03\textwidth,height=0.12\textwidth,trim=0 0 0 0,clip]{images/Image907_2/cbar.png}}
		\end{overpic}
		\\
		
		\multirow{2}{*}[+5ex]{\centering \rotatebox{90}{\scriptsize Coronal}} &
		\begin{overpic}[width=0.11\textwidth,angle=0,trim=50 5 50 20,clip]{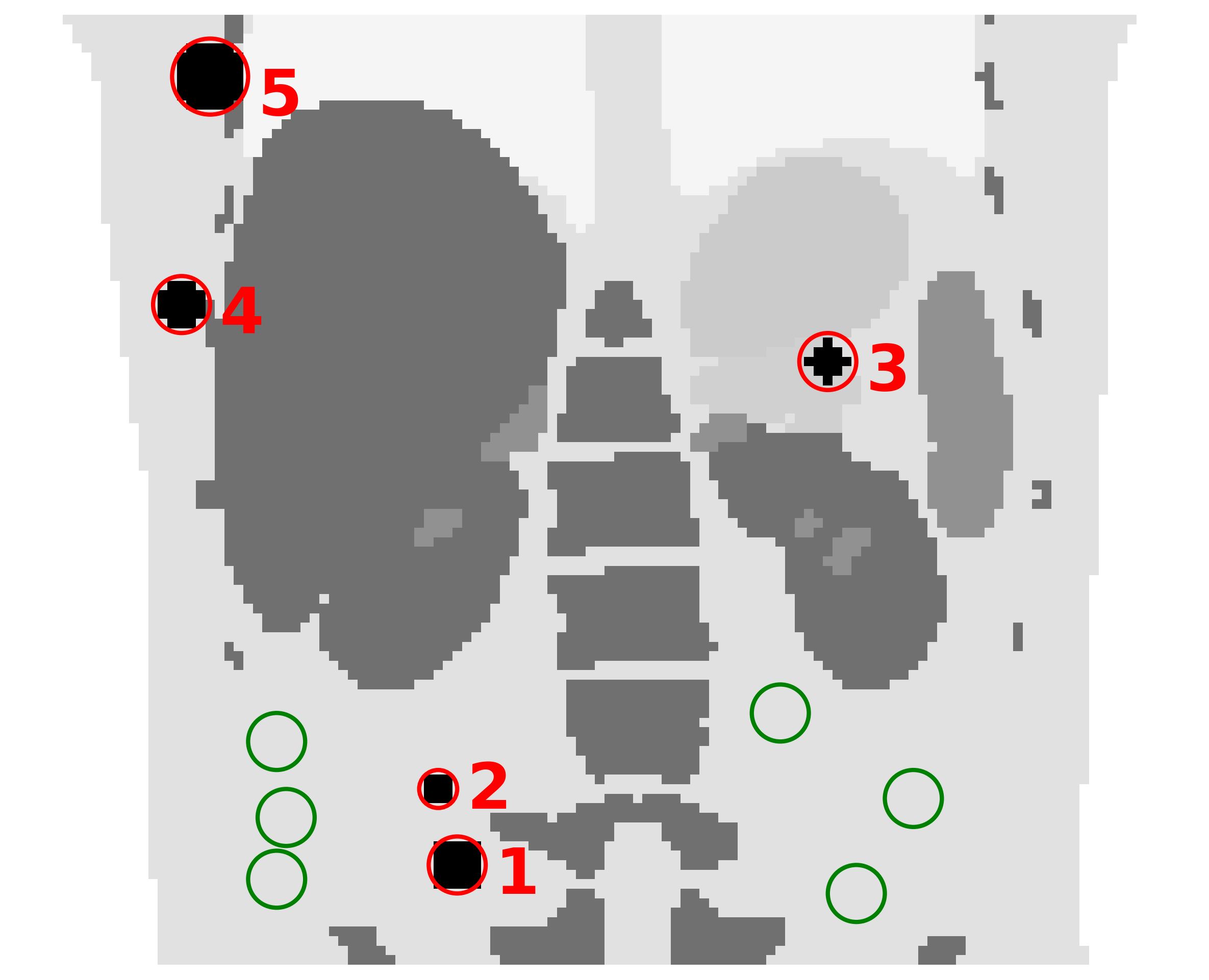}
		\end{overpic} &
		\begin{overpic}[width=0.11\textwidth,angle=0,trim=50 5 50 20,clip]{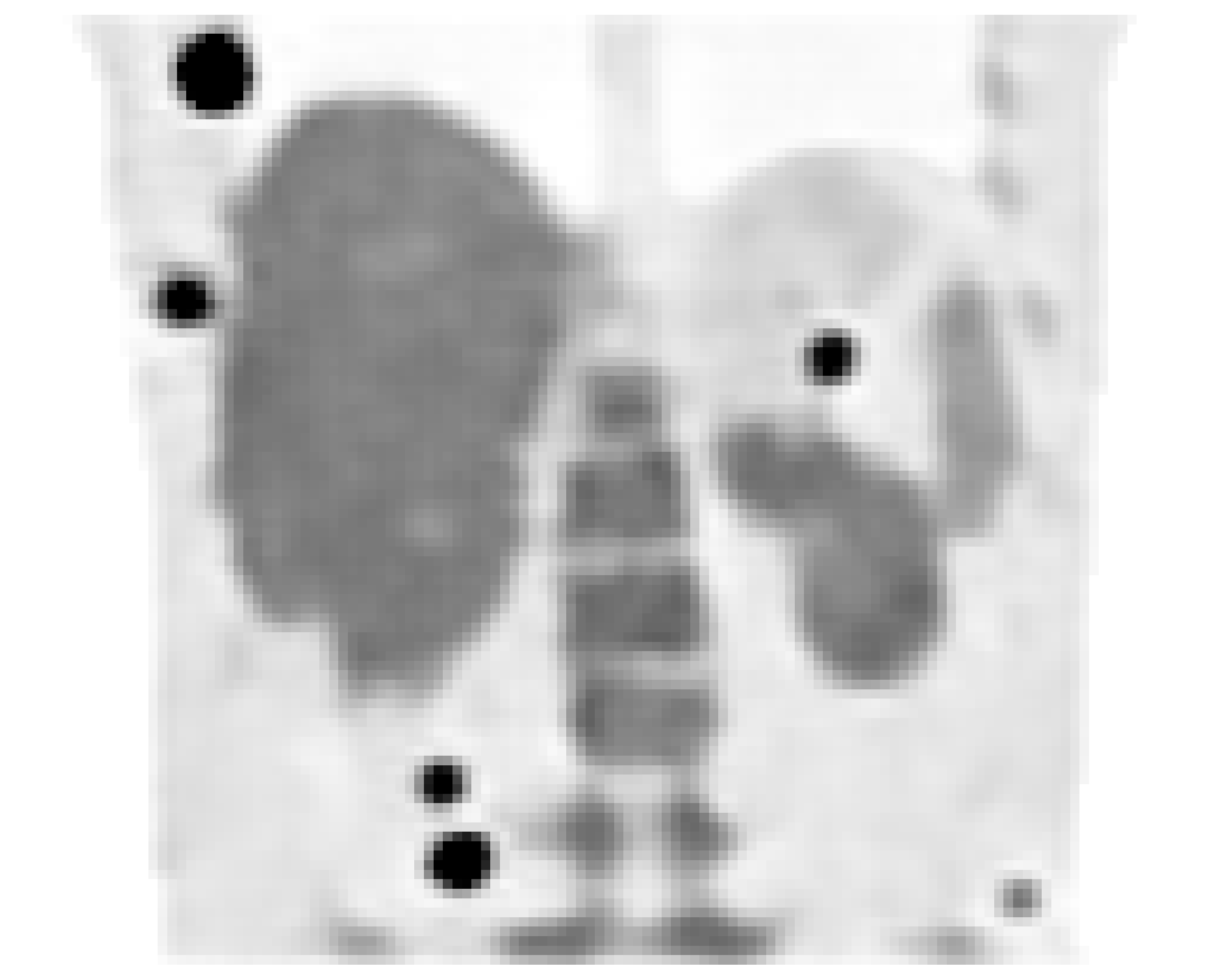}
		\end{overpic} &
		\begin{overpic}[width=0.11\textwidth,angle=0,trim=50 5 50 20,clip]{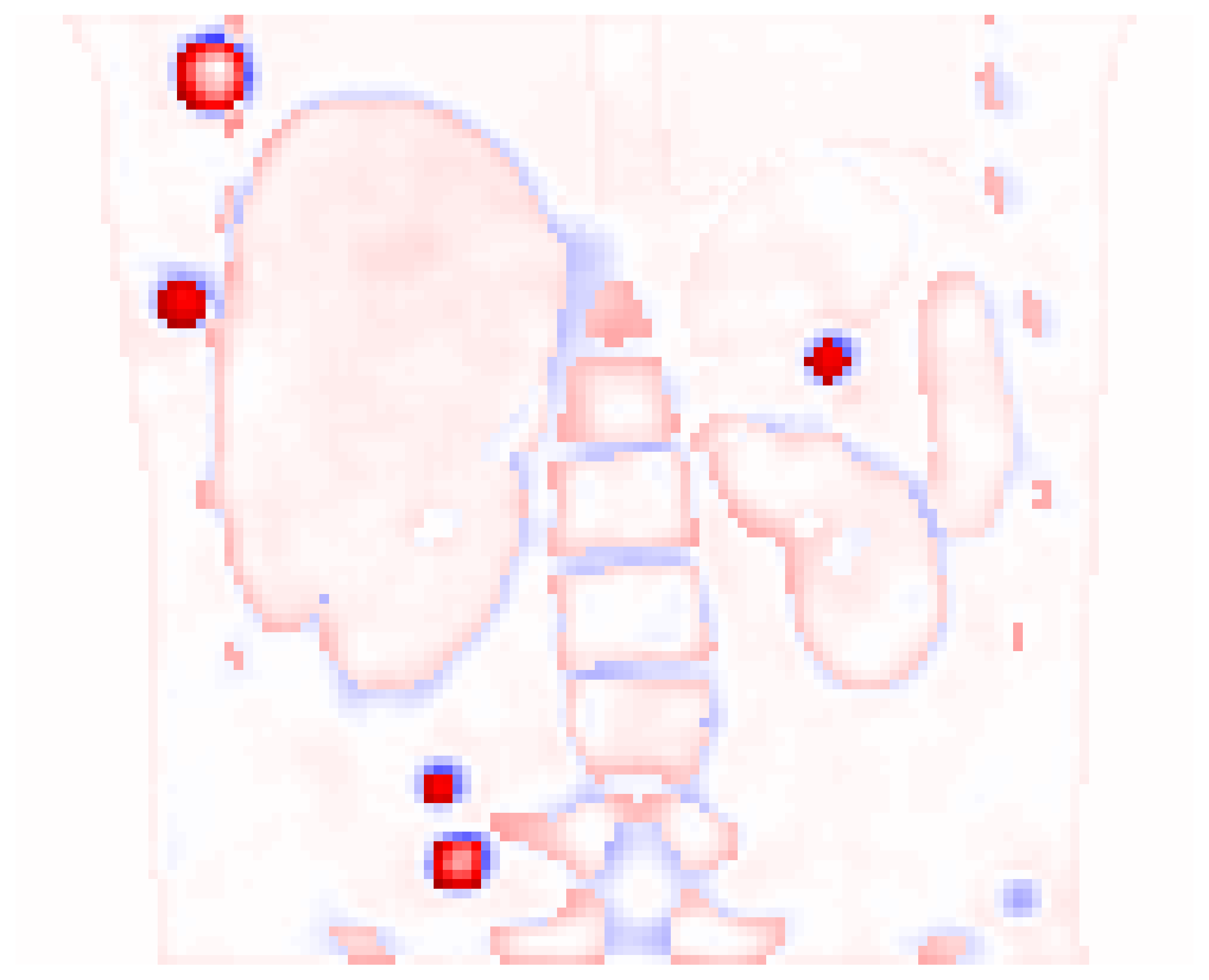}
		\end{overpic} &
		\begin{overpic}[width=0.11\textwidth,angle=0,trim=50 5 50 20,clip]{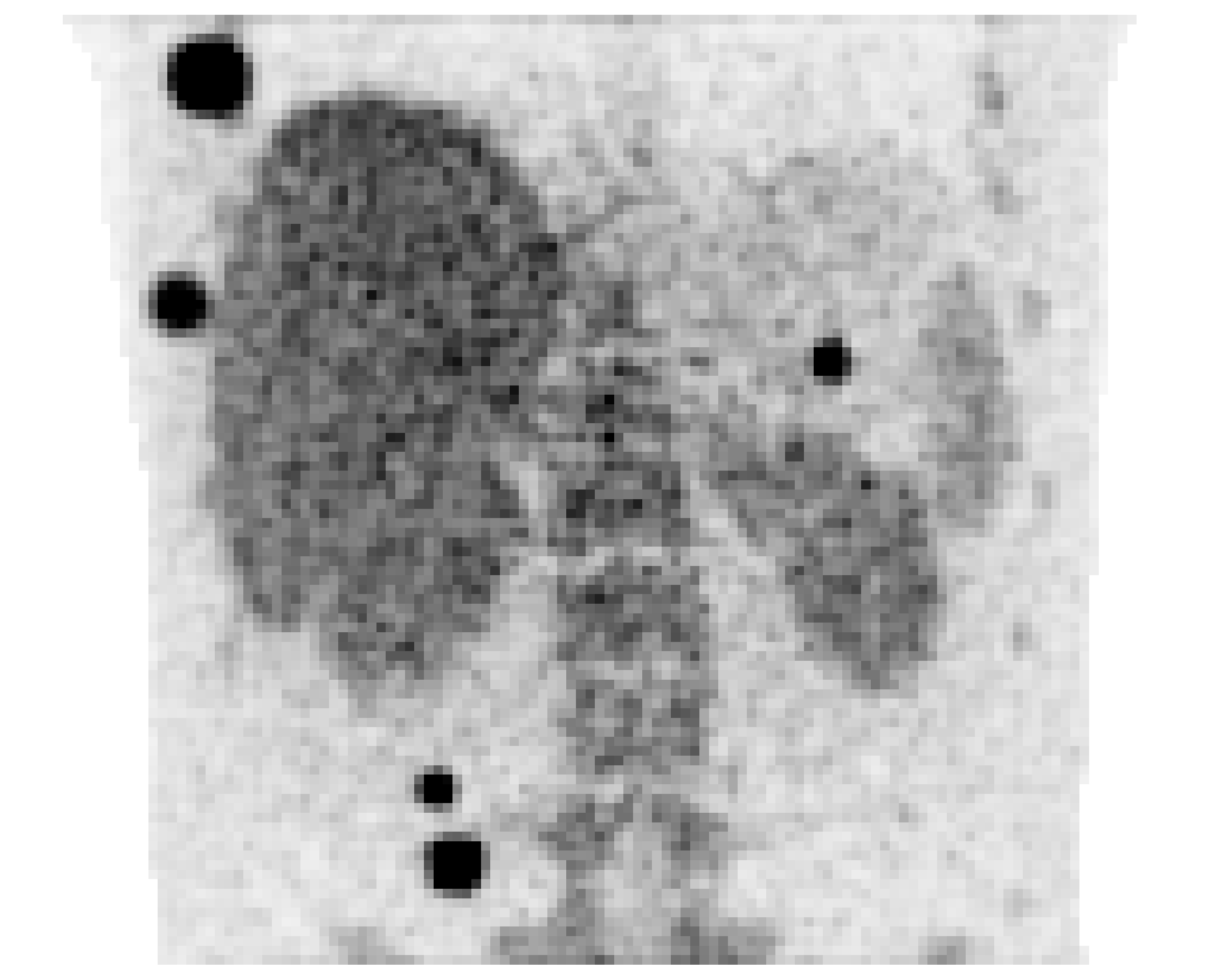}
		\end{overpic} &
		\begin{overpic}[width=0.11\textwidth,angle=0,trim=50 5 50 20,clip]{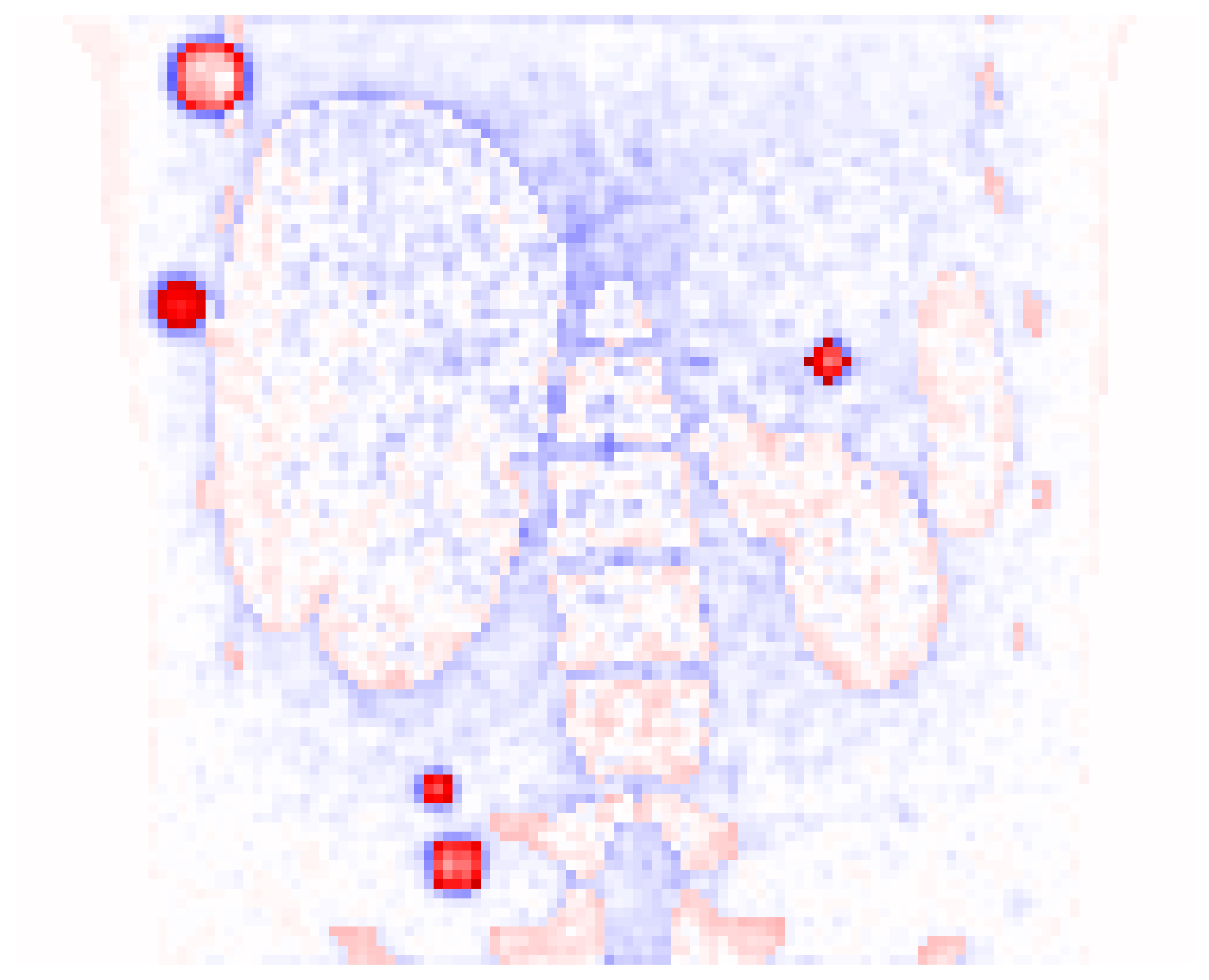}
		\end{overpic} &
		\begin{overpic}[width=0.11\textwidth,angle=0,trim=50 5 50 20,clip]{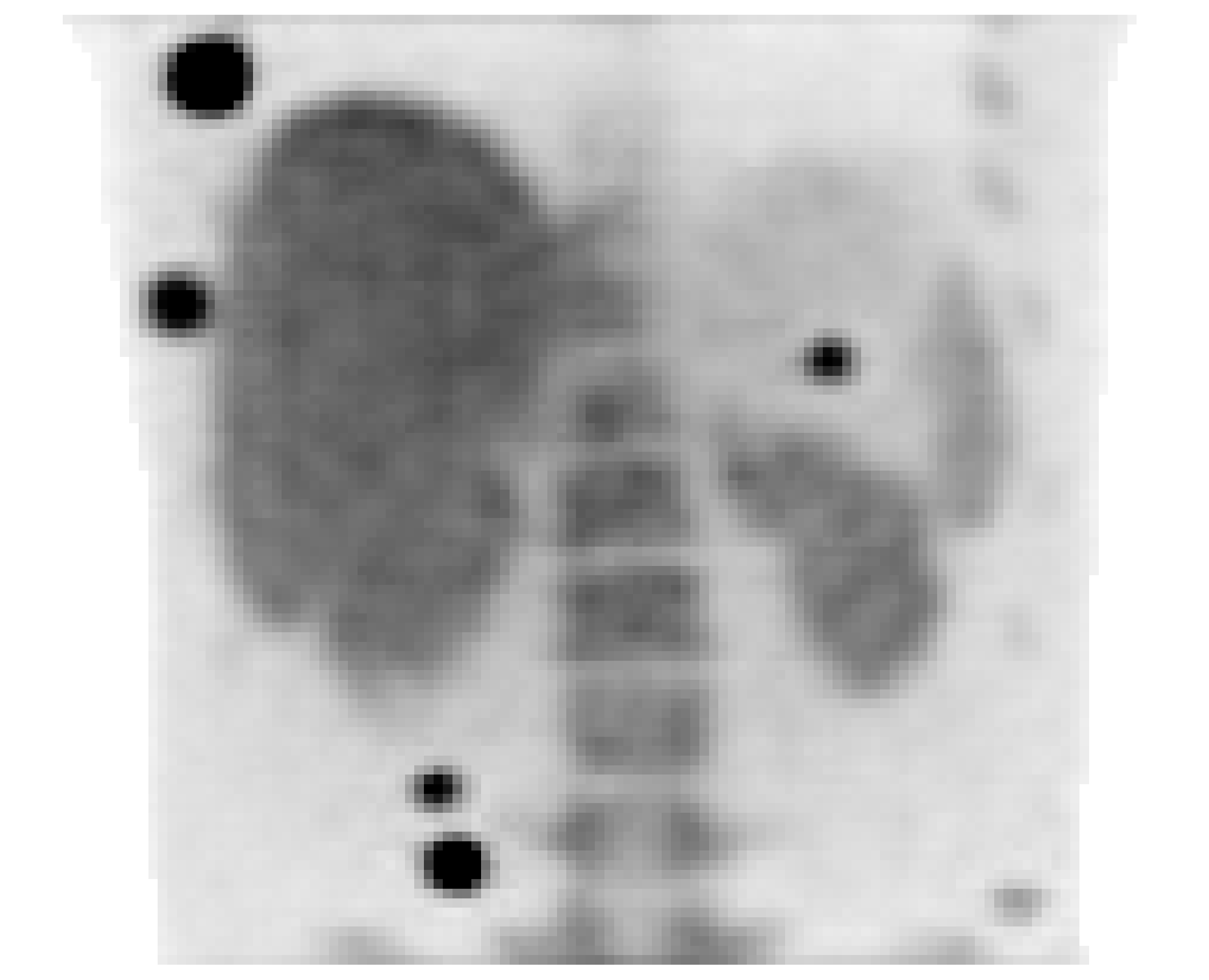}
		\end{overpic} &
		\begin{overpic}[width=0.11\textwidth,angle=0,trim=50 5 50 20,clip]{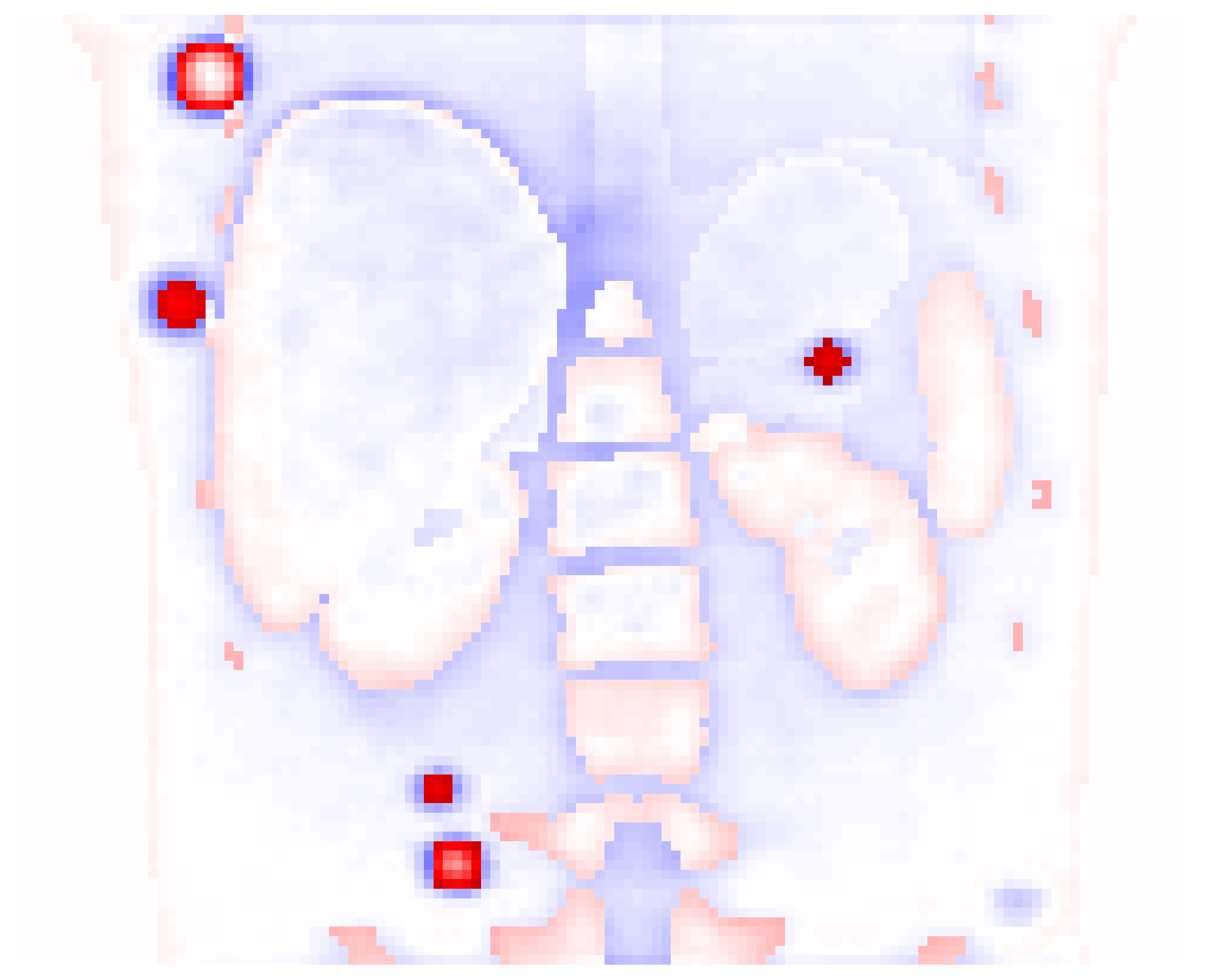}
		\end{overpic}\\
		
		\multirow{2}{*}[+6ex]{\centering \rotatebox{90}{\scriptsize Transverse}} &
		\begin{overpic}[width=0.11\textwidth,trim=6 3 6 3,clip]{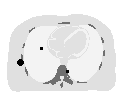}
		\end{overpic} &
		\begin{overpic}[width=0.11\textwidth,trim=6 3 6 3,clip]{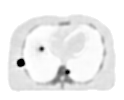}
		\end{overpic} &
		\begin{overpic}[width=0.11\textwidth,trim=6 3 6 3,clip]{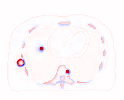}
		\end{overpic} &
		\begin{overpic}[width=0.11\textwidth,trim=6 3 6 3,clip]{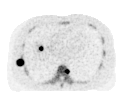}
		\end{overpic} &
		\begin{overpic}[width=0.11\textwidth,trim=6 3 6 3,clip]{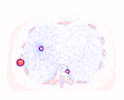}
		\end{overpic} &
		\begin{overpic}[width=0.11\textwidth,trim=6 3 6 3,clip]{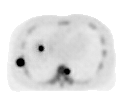}
		\end{overpic} &
		\begin{overpic}[width=0.11\textwidth,trim=6 3 6 3,clip]{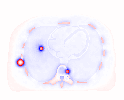}
		\end{overpic}\\
	\end{tabular}

	\caption{
		Experiment 2 (human-sized scanner)---Comparison of Direct3$\upgamma$, \ac{TOF} and \ac{DIP}-\ac{TOF} reconstructions (Phantom 1). The coronal view of the \ac{GT} emission image with overlaid red spherical \acp{ROI} of different radius (lesion 1: 6~mm; lesion 2: 4~mm; lesion 3: 6~mm; lesion 4: 6~mm; lesion 5: 8~mm)---background \acp{ROI} are shown as  6-mm radius green spheres.
	}
	\label{fig:combined_views_with_error_human}
\end{figure*}

A comparison of reconstruction methods is presented in Figure \ref{fig:combined_views_with_error_human}. The results suggest that Direct3$\upgamma$ consistently produces higher-quality images than both \ac{TOF} and \ac{DIP}-\ac{TOF}. Direct3$\upgamma$ provides clearer anatomical details and fewer errors compared to the \ac{GT}, particularly in the coronal and sagittal views, where contrast is better preserved and the cold regions appear more distinct, although we observe a contrast drop in the upper vertebra. While \ac{TOF} offers sharper images, the reconstructed images are noisier than Direct3$\upgamma{}$, and \ac{DIP}-\ac{TOF}, though effective in noise reduction, oversmooths important structures, leading to loss of details.

Figure~\ref{fig:ssim_psnr_comparison_human} shows the scatter plot of the distribution of \acs{SSIM} and \acs{PSNR} values for all methods across the five  \ac{XCAT} test phantoms. The Direct3$\upgamma$ method consistently achieved superior \acs{SSIM} values, ranging from 0.9399 to 0.9721, along with \acs{PSNR} values between 26.15 and 35.40. In contrast, the classical \ac{TOF} images exhibited lower \acs{SSIM} values between 0.8708 and 0.9475, and \acs{PSNR} values ranging from 25.74 to 34.37. A slight improvement in \ac{PSNR} is observed with \ac{DIP}-\ac{TOF} images, with values ranging from 26.53 to 34.82 as \ac{DIP} tends to reduce noise and smooth the reconstructed images. However, the \acs{SSIM} values for \ac{DIP}-\ac{TOF}, ranging between 0.8612 and 0.9390, remain slightly lower than  \ac{TOF} images, likely due to the over-smoothing effect of \ac{DIP}, which can reduce fine structural details. While \ac{DIP} improved the noise characteristics (as reflected in higher \acs{PSNR}), its impact on structural preservation was more modest, leading to only marginal improvements in \acs{SSIM} compared to classical \ac{TOF}.

\begin{figure}[htbp]
    \centering
    \begin{tikzpicture}
        \begin{axis}[
            scatter/classes={
                ThreeGamma={mark=square*,blue},
                TOF={mark=triangle*,red},
                DIPTOF={mark=diamond*,green}
            },
            xlabel={SSIM},
            ylabel={PSNR},
            legend style={font=\small, at={(0.2,0.99)}, anchor=north, legend columns=1}, 
            width=0.9\columnwidth,
            height=0.6\columnwidth,
            grid=major,
            scatter,
            only marks,
            label style={font=\small}, 
            tick label style={font=\small} 
        ]
	
	        \addplot[scatter, scatter src=explicit symbolic]
	        coordinates {
	            (0.9400, 26.15) [ThreeGamma]
	            (0.9399, 26.30) [ThreeGamma]
	            (0.9708, 35.40) [ThreeGamma]
	            (0.9721, 33.79) [ThreeGamma]
	            (0.9623, 33.43) [ThreeGamma]
	        };
	
	        \addplot[scatter, scatter src=explicit symbolic]
	        coordinates {
	            (0.8743, 25.74) [TOF]
	            (0.8708, 25.89) [TOF]
	            (0.9475, 34.37) [TOF]
	            (0.9464, 33.47) [TOF]
	            (0.9183, 29.88) [TOF]
	        };
	
	        \addplot[scatter, scatter src=explicit symbolic]
	        coordinates {
	            (0.8657, 26.53) [DIPTOF]
	            (0.8612, 26.88) [DIPTOF]
	            (0.9390, 34.82) [DIPTOF]
	            (0.9375, 34.23) [DIPTOF]
	            (0.9091, 30.12) [DIPTOF]
	        };
	
	        \legend{Direct3$\upgamma$, \ac{TOF}, \ac{DIP}-\ac{TOF} }
        \end{axis}
    \end{tikzpicture}
    \caption{Experiment 2 (human-sized scanner)---Scatter plot illustrating the \ac{SSIM} and \ac{PSNR} values for Direct3$\upgamma$,  \ac{TOF} and \ac{DIP}-\ac{TOF}, evaluated across the five human-sized phantoms (Figures~\ref{subfig:Phantom1}--\subref{subfig:Phantom5}).}
    \label{fig:ssim_psnr_comparison_human}
\end{figure}

Additionally, we we evaluated the detectability of the five lesions in Phantom 1 
(red spheres in Figure~\ref{fig:combined_views_with_error_human}, coronal view, \ac{GT})
with respect to background areas (green spheres in Figure~\ref{fig:combined_views_with_error_human}, Coronal view, \ac{GT}). Denoting $\mu_A$ and $\sigma_A$ the mean and \ac{STD} in area $A$ respectively, we used the following metrics: \ac{CRC}, \ac{CNR} and \ac{RC}, defined as
\begin{equation}\label{eq:cnr}
	\mathrm{CNR} = \frac{ \mu_{\mathrm{lesion}}  -  \mu_{\mathrm{background}} }{ \sigma_{\mathrm{bg}} } \,,
    \nonumber
\end{equation}
\begin{equation}
	\mathrm{CRC} = \frac{ \mu_{\mathrm{lesion}} }{ \mu_{\mathrm{background}} } - 1 \,
	\label{eq:crc}
    \nonumber
\end{equation}
\begin{equation}
	\mathrm{RC} = \frac{ \mu_{\mathrm{lesion}} }{ \mu_{\mathrm{lesion}}^{\star} }  
	\label{eq:rc}
    \nonumber
\end{equation}
where $\mu_{\mathrm{lesion}}^{\star}$ is the true mean activity in the lesion (from the phantom).

The results, summarized in Table~\ref{tab:lesion-metrics}, indicate that Direct3$\upgamma$ consistently outperforms \ac{TOF} and \ac{DIP}-\ac{TOF} across all metrics evaluated. 

\begin{table}[ht]
	\centering
	\caption{Experiment 2 (human-sized scanner)---Comparison of Direct3$\upgamma$, \ac{TOF} and \ac{DIP}-\ac{TOF} with different lesion metrics.}
	\label{tab:lesion-metrics}
	\begin{tabular}{l l  c c c}
		\hline
		\textbf{Lesion} & \textbf{Method} & \textbf{\acs{CNR}} & \textbf{\acs{CRC}} & \textbf{\acs{RC}} \\
		\hline
		\multirow{3}{*}{Lesion 1}
		& Direct3$\upgamma$ &  95.74 & 11.41 & 0.70 \\
		& TOF        & 24.92 & 7.82 & 0.50 \\
		& DIP-TOF    & 30.02 & 7.28 & 0.46 \\
		\hline
		\multirow{3}{*}{Lesion 2}
		& Direct3$\upgamma$  & 61.59 & 7.34 & 0.58 \\
		& TOF         & 18.66 & 5.86 & 0.47 \\
		& DIP-TOF     & 20.89 & 5.07 & 0.42 \\
		\hline
		\multirow{3}{*}{Lesion 3}
		& Direct3$\upgamma$  & 48.68 & 5.80 & 0.68 \\
		& TOF         & 13.61 & 4.27 & 0.53 \\
		& DIP-TOF     & 16.30 & 3.95 & 0.50 \\
		\hline
		\multirow{3}{*}{Lesion 4}
		& Direct3$\upgamma$  & 49.96 & 5.96 & 0.46 \\
		& TOF         & 16.77 & 5.26 & 0.41 \\
		& DIP-TOF     & 20.17 & 4.89 & 0.39 \\
		\hline
		\multirow{3}{*}{Lesion 5}
		& Direct3$\upgamma$  & 112.72 & 13.44 & 0.76 \\
		& TOF         & 33.82 & 10.62 & 0.62 \\
		& DIP-TOF     & 42.73 & 10.36 & 0.60 \\
		\hline
	\end{tabular}
\end{table}

\subsubsection{Mouse-sized Scanner}

\begin{figure*}[htbp]
	\centering
	\begin{tabular}{c cccccccc}
		& \scriptsize \acs{GT} & \scriptsize Direct3$\upgamma$ & \scriptsize Direct3$\upgamma$ Error & \scriptsize \ac{TOF} & \scriptsize \ac{TOF} Error & \scriptsize \ac{DIP}-\ac{TOF} & \scriptsize \ac{DIP}-\ac{TOF} Error \\
		
		\multirow{2}{*}[7ex]{\centering \rotatebox{90}{\scriptsize Saggital}} &  
		\begin{overpic}[width=0.09\textwidth,angle=90,trim=9 0 9 0,clip]{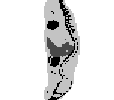}
		\end{overpic} &
		\begin{overpic}[width=0.09\textwidth,angle=90,trim=9 0 9 0,clip]{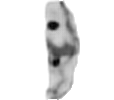}
		\end{overpic} &
		\begin{overpic}[width=0.09\textwidth,angle=90,trim=9 0 9 0,clip]{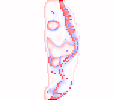}
			\put(105,0){\includegraphics[width=0.025\textwidth,height=0.1\textwidth,trim=0 0 0 0,clip]{images/Image907_2/cbar.png}}
		\end{overpic} &
		\begin{overpic}[width=0.09\textwidth,angle=90,trim=9 0 9 0,clip]{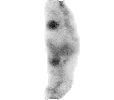}
		\end{overpic} &
		\begin{overpic}[width=0.09\textwidth,angle=90,trim=9 0 9 0,clip]{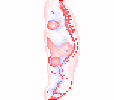}
			\put(105,0){\includegraphics[width=0.025\textwidth,height=0.1\textwidth,trim=0 0 0 0,clip]{images/Image907_2/cbar.png}}
		\end{overpic} &
		\begin{overpic}[width=0.09\textwidth,angle=90,trim=9 0 9 0,clip]{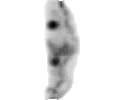}
		\end{overpic} &
		\begin{overpic}[width=0.09\textwidth,angle=90,trim=9 0 9 0,clip]{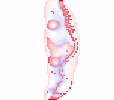}
			\put(105,0){\includegraphics[width=0.025\textwidth,height=0.1\textwidth,trim=0 0 0 0,clip]{images/Image907_2/cbar.png}}
		\end{overpic}
		\\
		
		\multirow{2}{*}[+6.5ex]{\centering \rotatebox{90}{\scriptsize Coronal}} &  
		\begin{overpic}[width=0.09\textwidth,angle=90,trim=9 0 9 0,clip]{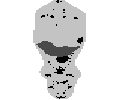}
		\end{overpic} &
		\begin{overpic}[width=0.09\textwidth,angle=90,trim=9 0 9 0,clip]{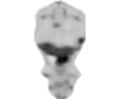}
		\end{overpic} &
		\begin{overpic}[width=0.09\textwidth,angle=90,trim=9 0 9 0,clip]{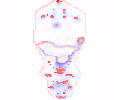}
		\end{overpic} &
		\begin{overpic}[width=0.09\textwidth,angle=90,trim=9 0 9 0,clip]{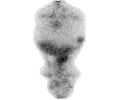}
		\end{overpic} &
		\begin{overpic}[width=0.09\textwidth,angle=90,trim=9 0 9 0,clip]{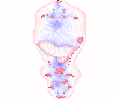}
		\end{overpic} &
		\begin{overpic}[width=0.09\textwidth,angle=90,trim=9 0 9 0,clip]{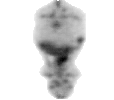}
		\end{overpic} &
		\begin{overpic}[width=0.09\textwidth,angle=90,trim=9 0 9 0,clip]{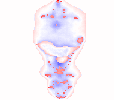}
		\end{overpic}\\
		
		\multirow{2}{*}[+6ex]{\centering \rotatebox{90}{\scriptsize Transverse}} &  
		\begin{overpic}[width=0.1\textwidth,angle=180,trim=20 20 20 20,clip]{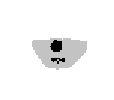}
		\end{overpic} &
		\begin{overpic}[width=0.1\textwidth,angle=180,trim=20 20 20 20,clip]{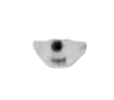}
		\end{overpic} &
		\begin{overpic}[width=0.1\textwidth,angle=180,trim=20 20 20 20,clip]{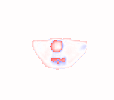}
		\end{overpic} &
		\begin{overpic}[width=0.1\textwidth,angle=180,trim=20 20 20 20,clip]{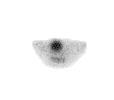}
		\end{overpic} &
		\begin{overpic}[width=0.1\textwidth,angle=180,trim=20 20 20 20,clip]{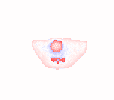}
		\end{overpic} &
		\begin{overpic}[width=0.1\textwidth,angle=180,trim=20 20 20 20,clip]{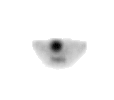}
		\end{overpic} &
		\begin{overpic}[width=0.1\textwidth,angle=180,trim=20 20 20 20,clip]{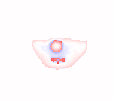}
		\end{overpic}\\
	\end{tabular}
	\caption{Experiment 2 (mouse-sized scanner)---Comparison of Direct3$\upgamma$, \ac{TOF} and \ac{DIP}-\ac{PET} reconstructions.}
	\label{fig:combined_views_with_error_mouse}
\end{figure*}

A comparison between all reconstruction methods are presented in Figure~\ref{fig:combined_views_with_error_mouse}. It shows that the Direct3$\upgamma$ reconstruction offers a smoother image with fewer errors, though some finer details are missing compared to the \ac{GT}. The \ac{TOF} image suffers from significant noise, while the \ac{DIP}-\ac{TOF} image, though effective at noise reduction, oversmooths the image. In the error maps, the Direct3$\upgamma$ method exhibits fewer and less intense errors compared to \ac{TOF} and \ac{DIP}-\ac{TOF}, indicating a better balance between reducing error and maintaining image clarity.

Figure~\ref{fig:ssim_psnr_comparison_small_scanner} shows a scatter plot for quantitative metrics for the five mouse-sized \ac{XCAT} test phantoms. The results show Direct3$\upgamma$ yielding higher \acs{SSIM} values, ranging from 0.9302 to 0.9354, and \acs{PSNR} values between 25.31 and 26.77. In comparison, the \ac{TOF} images produced \acs{SSIM} values ranging from 0.8881 to 0.8913, and \acs{PSNR} values between 22.36 and 23.03. \Ac{DIP}-\ac{TOF} displayed a slight improvement in terms of \ac{PSNR}, with values ranging from 22.90 to 23.50. However, the \ac{DIP}-\ac{TOF} \acs{SSIM} values ranging from 0.8825 to 0.8895, remained slightly lower than  \ac{TOF} due to over-smoothing effects.

\begin{figure}[htbp]
    \centering
    \begin{tikzpicture}
        \begin{axis}[
            scatter/classes={
                ThreeGamma={mark=square*,blue},
                TOF={mark=triangle*,red},
                DIPTOF={mark=diamond*,green}
            },
            xlabel={SSIM},
            ylabel={PSNR},
            legend style={font=\small, at={(0.2,0.99)}, anchor=north, legend columns=1}, 
            width=0.9\columnwidth,
            height=0.6\columnwidth,
            grid=major,
            scatter,
            only marks,
            label style={font=\small}, 
            tick label style={font=\small} 
        ]

        \addplot[scatter, scatter src=explicit symbolic]
        coordinates {
            (0.9354, 26.77) [ThreeGamma]
            (0.9332, 26.21) [ThreeGamma]
            (0.9329, 25.77) [ThreeGamma]
            (0.9309, 25.31) [ThreeGamma]
            (0.9302, 25.39) [ThreeGamma]
        };

        \addplot[scatter, scatter src=explicit symbolic]
        coordinates {
            (0.8913, 22.36) [TOF]
            (0.8912, 22.72) [TOF]
            (0.8909, 23.03) [TOF]
            (0.8901, 22.87) [TOF]
            (0.8881, 22.62) [TOF]
        };

        \addplot[scatter, scatter src=explicit symbolic]
        coordinates {
            (0.8850, 22.90) [DIPTOF]
            (0.8842, 23.15) [DIPTOF]
            (0.8895, 23.50) [DIPTOF]
            (0.8880, 23.30) [DIPTOF]
            (0.8825, 23.00) [DIPTOF]
        };

        \legend{Direct3$\upgamma$, \ac{TOF}, \ac{DIP}-\ac{TOF}}
        \end{axis}
    \end{tikzpicture}
    \caption{Experiment 2 (mouse-sized scanner)---Scatter plot illustrating the \ac{SSIM} and \ac{PSNR} values for Direct3$\upgamma$, \ac{TOF} and \ac{DIP}-\ac{TOF}, evaluated across the five mouse-sized phantoms (Figures~\ref{subfig:Phantom1m}--\subref{subfig:Phantom5m}).}
    \label{fig:ssim_psnr_comparison_small_scanner}
\end{figure}


%% file: content/discussion.tex
\section{Discussion}\label{sec:discussion}

This study introduces a comprehensive approach to \ac{3g} \ac{PET} imaging using \textsuperscript{44}Sc, addressing key challenges in the determination of the sequence of the photon interaction and the estimation of emission points. Our  proposed \acf{MIN} demonstrated better accuracy in determining photon interaction sequences compared to both physical ($\rmd\phi$-criterion) and classical \acf{FCNN} methods, particularly when complexity increases and the number of interactions of the prompt gamma is greater than 4. This improvement in sequence determination is crucial for the determination of Compton cone parameters, thus accurate image reconstruction in \ac{3g} \ac{PET}. However, \ac{MIN} accuracy declines for $N > 5$ interactions due to the increasing combinatorial complexity ($N!$). Additionally, \ac{MIN} occasionally generates nonadmissible graphs that violate physical constraints. Additionally, reliance on \ac{MC} training data introduces potential biases, as detector effects such as energy resolution, spatial uncertainty, and Doppler broadening may affect real-world performance \cite{feng2021influence}. \Ac{MIN} is also sensitive to detector noise, struggling with sequences where consecutive interactions deposit similar energy, as it lacks explicit enforcement of Compton kinematics. Furthermore, \ac{MIN} assumes isolated prompt gammas, making it ineffective in high-emission environments where multiple photons interact within short time windows. Extending its capabilities to handle multiphoton events is essential for broader applicability in high-count-rate imaging.

We introduced a novel \acf{DREP} module and implemented a nonsymmetric Gaussian function in our \ac{3g} histogrammer, both of which have demonstrated significant potential to improve the accuracy of emission event distribution modeling. The \ac{DREP} module specifically addresses two key types of measurement uncertainties: energy resolution errors and spatial position uncertainties. These uncertainties, inherent in \ac{PET} detectors, can significantly impact the accuracy of Compton cone reconstruction and consequently the precision of emission point estimation. Our \ac{3g} histogrammer utilizes a nonsymmetric Gaussian kernel to more realistically represent the probability distribution of emission points along the \ac{LOR}. This method accounts for the asymmetric nature of error propagation from the Compton cone to the \ac{LOR}, a factor often overlooked in conventional symmetric models.

Furthermore, we introduced a custom attenuation correction technique specifically designed for \textsuperscript{44}Sc-based \ac{3g} events. Unlike the standard attenuation correction applied in conventional \ac{PET}, which deals only with the back-to-back annihilation photons, our approach accounts for the attenuation effects of both 511-keV annihilation photons and the 1.157-MeV prompt gamma ray emitted during \textsuperscript{44}Sc decay. This dual-energy correction is essential, especially when considering larger or denser regions of interest, where photon attenuation is more prominent. By applying this correction step to the histoimages data prior to a \ac{NN} for image reconstruction, we provide the network with more accurate inputs. This ensures that the \ac{NN} focuses on mapping the true activity distribution, rather than compensating for physical effects such as attenuation.

To address issues of histoimages coming from incorrect Compton cone estimation, and the presence of multiple estimated points on the same \ac{LOR}, as well as low contrast due to the blurring effects introduced by \ac{DREP} on \acp{LOR}, we proposed a \ac{3D} U-Net trained to map histoimages to real emission maps. Initial training of the network with a single supervised loss resulted in blurred reconstructions. Therefore, we improved the training by integrating an adversarial loss component, which improved the preservation of fine details in the reconstructed images. Although adversarial loss helps preserve fine structural details, we recognize that such techniques carry the risk of introducing hallucinated image features \cite{cohen2018distribution}. In our experiments (with known \ac{GT} phantoms), no such spurious structures were found; all the enhanced details in the Direct3$\upgamma$ reconstructions corresponded to the actual features of the target images. This outcome reflects the careful balance in our training between adversarial and supervised losses. We highlight this point to warn that, as with any \ac{GAN}-based method, thorough validation is needed to ensure that no clinically misleading artifacts are produced. However, the training of the U-Net depends on the statistics. This can be addressed by transfer learning as proposed by \citeauthor{whiteley2020fastpet}~\cite{whiteley2020fastpet}. 

In both human-sized and small animal scanners, the Direct3$\upgamma$ method consistently shows better performance compared to both the \ac{TOF} images and  \ac{DIP}-\ac{TOF} images, particularly in terms of structural similarity and image clarity. This improvement can be attributed to the improved capabilities of the \ac{3g} system, in conjunction with the proposed \ac{DREP}, which allows for a more precise generation of histoimages and a better localization of the emission source. Furthermore, the integration of \acp{NN}, potentially augmented through adversarial training, improved reconstruction accuracy by improving noise reduction and structural fidelity. The \ac{DIP}-\ac{TOF} method offered a modest improvement in \acs{PSNR}, due to its noise reduction capabilities, but the associated decline in \acs{SSIM} can be related to over-smoothing, which reduced the retention of structural details. In small animal scanners, the slight drop in \acs{SSIM} and \acs{PSNR} is likely due to the reduction in voxel size used for reconstruction (from 3\texttimes{}3\texttimes{}3~mm\textsuperscript{3} in large scanners to 0.8\texttimes{}0.8\texttimes{}0.8~mm\textsuperscript{3}). This smaller voxel size leads to blurriness and loss of detail, which is due to the positron range effect, where positrons travel a short distance before annihilation, making it harder to maintain fine image resolution in smaller-scale scans.

%% file: content/conclusion.tex
\section{Conclusion}\label{sec:conclusion}

This study presents Direct3$\upgamma$, a novel approach for reconstructing \ac{3g} \ac{PET} images using \textsuperscript{44}Sc, addressing critical challenges in photon interaction sequencing and emission modeling. The proposed \ac{MIN} method significantly improves the photon ordering accuracy compared to existing approaches, particularly for complex interactions involving more than four events. Additionally, incorporating the \ac{DREP} and nonsymmetric Gaussian modeling effectively accounts for measurement uncertainties, improving emission event localization.

Our dual-energy attenuation correction technique further enhances quantitative accuracy by accounting for both 511-keV annihilation photons and the 1.157-MeV prompt gamma. Integration of a \ac{NN} for translating histoimages into high-resolution reconstructed images significantly improves detail retention and reduces noise compared to traditional \ac{TOF} \ac{PET} with \ac{MLEM}. Future research will address the challenge posed by \ac{PR} effects to further enhance small-scale imaging applications.

%% file: content/appendix.tex

\section{Edge Classification}\label{sec:classification}

In \cite{Interaction_networks} the graph is built using two separate matrices that define how messages are passed between nodes. The sender and receiver matrices, respectfully denoted $\boldR_\rms\in \R^{N\times N_\rme}$ and  $\boldR_\rmr\in \R^{N\times N_\rme}$, are defined as
\begin{equation}\label{eq:senders}
	[\boldR_\rms]_{n,l} = \left \{
	\begin{array}{ll}
		1 & \text{if edge $l$ departs from node $n$} \\
		0 & \text{otherwise} 
	\end{array}	
	\right. \notag
\end{equation}
and 
\begin{equation}\label{eq:receivers.}
	[\boldR_\rmr]_{n,l} = \left \{
	\begin{array}{ll}
		1 & \text{if edge $l$ arrives at node $n$} \\
		0 & \text{otherwise} 
	\end{array}	
	\right. \, \notag
\end{equation}
For example the sender and receiver matrices corresponding to the graph in Figure~\ref{subfig:I_Graph} are
\begin{equation}
	\boldR_\rms = \kbordermatrix{
		& \bolde_1 & \bolde_2 & \bolde_3 & \bolde_4 & \bolde_5 & \bolde_6 \\
		\boldo_1 & 1 & 0 & 1 & 0 & 0 & 0 \\
		\boldo_2 & 0 & 1 & 0 & 0 & 1 & 0 \\
		\boldo_3 & 0 & 0 & 0 & 1 & 0 & 1
	} \notag
	\label{eq:send_matrix2}
\end{equation}
and
\begin{equation}
	\boldR_\rmr = \kbordermatrix{
		& \bolde_1 & \bolde_2 & \bolde_3 & \bolde_4 & \bolde_5 & \bolde_6 \\
		\boldo_1 & 0 & 1 & 0 & 1 & 0 & 0 \\
		\boldo_2 & 1 & 0 & 0 & 0 & 0 & 1 \\
		\boldo_3 & 0 & 0 & 1 & 0 & 1 & 0
	} \notag
	\label{eq:reciving_matrix2}
\end{equation}
which should be read for example (first column of $\boldR_\rms$ and $\boldR_\rme$) as ``$\bolde_1$ \emph{sends information from} $\boldo_1$ \emph{to} $\boldo_2$''.

\begin{figure}[!h]
	\centering
	\begin{subfigure}[b]{0.48\columnwidth}
		\centering
		\input{./tikz/possibilities}
		\caption{Input graph}\label{subfig:I_Graph}
	\end{subfigure}%
	\begin{subfigure}[b]{0.48\columnwidth}
		\centering
		\input{./tikz/gt_graph}
		\caption{Possible output graph} \label{subfig:O_Graph}
	\end{subfigure}
	\caption{\subref{subfig:I_Graph} Fully connected graph used as an input for the \ac{GNN}; \subref{subfig:O_Graph} possible output of the \ac{GNN}. 
	}
	\label{Edge_classification}
\end{figure}

The first step in the design of an \ac{IN} is to create a message matrix $\boldM$ defined as 
\begin{equation}
	\boldM = [\boldR_\rms \transp  \boldO, \boldR_\rmr \transp  \boldO ] \in \R^{N_\rme \times 2N_\rmf} \,  .\nonumber
\end{equation}
where $\boldO = [\boldo_1\transp , \dots ,\boldo_N\transp ]\transp \in \R^{N\times N_\rmf}$ and $[\cdot,\cdot]$ denotes the horizontal concatenation of two matrices  with the same number of rows, the total number of directed edges is $N_\rme = N(N-1)$. Each row in $\boldM$ corresponds to an edge and contains $2N_\rmf$ features.

In a second step we encode the information carried in each edge into an \emph{effect vector}  of dimension $N_{\mathrm{eff}}$ via a \emph{relation-centric} \ac{NN} $f_\rmR:\mathbb{R}^{2 N_\rmf}\mapsto\mathbb{R}^{N_{\mathrm{eff}}}$ (we used $N_{\mathrm{eff}}=50$ as suggested in \citeauthor{Interaction_networks}~\cite{Interaction_networks}) which maps each row of $\boldM$ to an  $N_{\mathrm{eff}}$-dimensional row vector. By repeating the operation on each row, $f_\rmR$ maps $\boldM$ to an $N_\rme\times N_{\mathrm{eff}}$ matrix denoted $\boldE$ referred to as the \emph{effect matrix}. Each row in $\boldE$ represents a latent vector of effect for each node on its neighbor. 

In a third step the cumulative effect of interactions received by each node is stored in a $N\times N_{\mathrm{eff}}$  matrix $\bar{\boldE}$ defined as
\begin{equation}
	\bar{\boldE} = \boldR_\rmr \boldE \, \notag ,
\end{equation}
and we define the $N \times( N_\rmf + N_{\mathrm{eff}} )$ matrix 
\begin{equation}
	\boldC = [ \boldO , \bar{\boldE}  ]  \, \notag .
\end{equation}
Each row in $\boldC$ contains the node's features combined with the interaction effects. An \emph{object-centric} \ac{NN} $f_\rmO$ maps the $N_\rmf + N_\mathrm{eff}$ features for each of the $N$ nodes to a 10-dimensional vector, resulting in a $N\times10$ matrix which is then flattened into a $10N$-dimensional column vector. Finally, we used an \emph{edge} \ac{NN} $f_\rmE \colon \mathbb{R}^{10N}\to \mathbb{R}^{N_\rme}$ which maps the new node features vector  to a vector of $N_{\rme}$  weights, 
each weight corresponding to an edge. We used a sigmoid activation function in order to give a score between 0 and 1 for each edge.

We thus defined a mapping $\boldF_{\boldvartheta} \colon \R^{N\times N_\rmf}  \to [0,1]^{N_\rme}$ where the parameter $\boldvartheta$ regroups those of $f_\rmR$, $f_\rmO$ and $f_\rmE$.

\section{Propagation of Energy and Spatial Uncertainties Along the LOR}\label{sec:uncertainty}

To address the spatial and energy uncertainties that cause blurring along the \ac{LOR} in \ac{3g} \ac{PET} imaging, we have developed a \ac{DREP} module to determine $\sigma_{\pm}$ in Equation~\eqref{eq:sigma_uniform}. This module accounts for both energy and spatial measurement uncertainties in the estimation of the Compton scattering angle and its propagation to the \ac{LOR}.

\subsection{Propagation of Energy Measurement Uncertainty}

We consider \eqref{eq:kn} with $k=1$ and $\theta_{\rmc} = \theta_1^{\mathrm{kin}}$. The 
To find the uncertainty in the scattering angle $\theta_\rmc$ due to the uncertainty in the measured deposited energy $E_1 = \calE_0 - \calE_1$, we first take the derivative of $\cos( \theta_\rmc)$ with respect to $\calE_1$:
\begin{equation}
	\frac{\rmd\cos(\theta_\rmc) }{\rmd \calE_1} = \frac{m_\rme c^2}{\calE_1^2}   \, . \notag
\end{equation}
The uncertainty in $\cos(\theta_\rmc)$ is then given by:
\begin{align}
	\Delta \cos(\theta_{\rmc}) & {} = \frac{\rmd\cos(\theta_\rmc)}{\rmd \calE_1}  \Delta \calE_1 \notag \\
	& {} =   \frac{m_\rme c^2}{\calE_1^2}  \Delta \calE_1  \, . \label{eq:delta}
\end{align}
Finally, we convert this to the uncertainty in $\theta_\rmc$ due to energy measurement, denoted $\Delta \theta_{\rmc}^\mathrm{energy}$:
\begin{align}
	\Delta \theta_{\rmc}^\mathrm{energy} & {} = -\frac{\Delta \cos(\theta_{\rmc})}{\sin(\theta_\rmc)}  \notag\\
	& {} =  -\frac{ \frac{m_\rme c^2}{\calE_1^2}  \Delta \calE_1}{\sin(\theta_\rmc)} \, \notag.
\end{align}
This equation shows how the error in the measured photon energy propagates to the uncertainty in the scattering angle.

\subsection{Mixing Energy and Spatial Measurement Uncertainties}

In a study on the detector resolution using the \acs{XEMIS}-1 \cite{manzano2016optimization} and \acs{XEMIS}-2 \cite{giovagnoli2020image} systems, it was shown that the spatial resolution contribution to the scatter angle error is almost constant and is approximately $\Delta \theta_{\rmc}^\mathrm{spatial}= 1.2^\rmo $.

Inspired from \cite{yoshida2020whole}, we consider variations in the angle $\theta_\rmc$ due to energy and spatial uncertainties:
\begin{align}
    \theta_{\rmc,\mathrm{energy}}^\pm &= \theta_\rmc \pm \Delta \theta_{\rmc}^\mathrm{energy}, \quad
    \theta_{\rmc,\mathrm{spatial}}^\pm &= \theta_\rmc \pm \Delta \theta_{\rmc}^\mathrm{spatial} \notag
\end{align}
These variations influence the intersection points of the Compton cone with the \ac{LOR}, resulting in the coordinates along the \ac{LOR} $(t_{\mathrm{energy}}^+, t_{\mathrm{energy}}^-, t_{\mathrm{spatial}}^+, t_{\mathrm{spatial}}^-)$. We define the blurring on the \ac{LOR} for each uncertainty as follows:
\begin{align}
    \sigma_{\mathrm{energy}}^\pm & = \left|t_{\boldp} - t_{\mathrm{energy}}^\pm\right|, \quad \sigma_{\mathrm{spatial}}^\pm = \left|t_{\boldp} - t_{\mathrm{spatial}}^\pm\right| \notag
\end{align}
where $t_{\boldp}$ is the estimated  emission position along the \ac{LOR} obtained by intersection with the Compton cone of angle $\theta_\rmc$.

Finally, we combine these uncertainties using the \ac{RSS} method to derive $\sigma_+$ and $\sigma_-$,
\begin{equation}\label{eq:sigma+-}
    \sigma_\pm = \sqrt{(\sigma_{\mathrm{energy}}^\pm)^2 + (\sigma_{\mathrm{spatial}}^\pm)^2} \, ,
\end{equation}
which are use to define the asymmetrical kernel \eqref{eq:sigma_uniform}.

%% file: tikz/possibilities.tex
\begin{tikzpicture}[scale=0.8, transform shape]
	\scriptsize
	\newlength{\dimtriangle}
	\setlength{\dimtriangle}{35pt}
	\newlength{\dimnode}
	\setlength{\dimnode}{17pt}
	
	\node(o1) at (90:\dimtriangle) [circle,inner sep=0pt,minimum size=\dimnode,draw=c1,fill=c11] {$\boldo_1$} ; 
	\node(o2) at (210:\dimtriangle) [circle,inner sep=0pt,minimum size=\dimnode,draw=c2,fill=c22] {$\boldo_2$} ; 
	\node(o3) at (-30:\dimtriangle) [circle,inner sep=0pt,minimum size=\dimnode,draw=c3,fill=c33] {$\boldo_3$} ; 
	
	\draw  [->,>=latex, line width=0.5pt,draw,dashed]  (o2.10) to node [above,midway] {$\bolde_5$} (o3.170) ;
	\draw  [->,>=latex, line width=0.5pt,draw,dashed]  (o3.190) to node [below,midway] {$\bolde_6$} (o2.-10) ;
	
	\draw  [->,>=latex, line width=0.5pt,draw,dashed]  (o1.230) to node [above,midway,sloped] {$\bolde_1$} (o2.70) ;
	\draw  [->,>=latex, line width=0.5pt,draw,dashed]  (o2.50) to node [below,midway,sloped] {$\bolde_2$} (o1.250) ;
	
	\draw  [->,>=latex, line width=0.5pt,draw,dashed]  (o1.-70) to node [below,midway,sloped] {$\bolde_3$} (o3.130) ;
	\draw  [->,>=latex, line width=0.5pt,draw,dashed]  (o3.110) to node [above,midway,sloped] {$\bolde_4$} (o1.-50) ;
	
\end{tikzpicture}	

%% file: tikz/gt_graph.tex
\begin{tikzpicture}[scale=0.8, transform shape]
	\scriptsize
	\setlength{\dimtriangle}{35pt}
	
	\setlength{\dimnode}{17pt}
	
	\node(o1) at (90:\dimtriangle) [circle,inner sep=0pt,minimum size=\dimnode,draw=c1,fill=c11] {$\boldo_1$} ; 
	\node(o2) at (210:\dimtriangle) [circle,inner sep=0pt,minimum size=\dimnode,draw=c2,fill=c22] {$\boldo_2$} ; 
	\node(o3) at (-30:\dimtriangle) [circle,inner sep=0pt,minimum size=\dimnode,draw=c3,fill=c33] {$\boldo_3$} ;

	\draw  [->,>=latex, line width=0.5pt,draw]  (o3) to node [below,midway] {$\bolde_6$} (o2) ;
	\draw  [->,>=latex, line width=0.5pt,draw]  (o1) to node [below,midway,sloped] {$\bolde_3$} (o3) ;

\end{tikzpicture}	